\documentclass[superscriptaddress,nofootinbib,11pt]{revtex4-1}

\usepackage{graphicx,amsmath,array,verbatim,setspace}
\usepackage{hyperref}       
\usepackage{url}            
\usepackage{amsfonts}       
\usepackage[italicdiff]{physics}
\usepackage{color}
\usepackage{amssymb,amsthm}
\usepackage{here}

\newcommand{\be}{\begin{eqnarray}}
\newcommand{\ee}{\end{eqnarray}}
\newcommand{\nn}{\nonumber}
\newcommand{\nl}{\nonumber \\}
\newcommand{\pd}{\partial}

\graphicspath{{./figures/}}

\begin{document}

\title{On exact-WKB analysis, resurgent structure, and quantization conditions}

\author{Naohisa Sueishi}
\email{sueishi@eken.phys.nagoya-u.ac.jp}
\address{Department of Physics, Nagoya University, Nagoya 464-8602, Japan}
 
\author{Syo Kamata}
\email{skamata11phys@gmail.com}
\address{College of Physics and Communication Electronics, Jiangxi Normal University, Nanchang 330022,
China}

\author{Tatsuhiro Misumi}
\email{misumi@phys.akita-u.ac.jp}
\address{Department of Mathematical Science, Akita University,  Akita 010-8502, Japan}
\address{Department of Physics, Keio University, Kanagawa 223-8521, Japan}

\author{Mithat \"{U}nsal}
\email{unsal.mithat@gmail.com}
\address{Department of Physics, North Carolina State University, Raleigh, NC 27607, USA}

\begin{abstract}
  
   There are two well-known approaches to studying nonperturbative aspects of quantum mechanical systems: saddle point 
   analysis  of  the partition functions in Euclidean 
   path integral formulation and  the exact-WKB analysis based on the wave functions in the Schr\"{o}dinger equation.  In this work, based on the quantization conditions obtained from the exact-WKB  method, 
   we determine the relations between the two formalism  and in particular show how the two Stokes phenomena are connected to each other: 
    the Stokes phenomenon leading to the ambiguous contribution of different sectors of the path integral formulation corresponds to the change of the ``topology'' of the Stoke curves in the exact-WKB analysis. 
        We also clarify the equivalence of different quantization conditions including Bohr-Sommerfeld, path integral and Gutzwiller's ones. 
In particular,  by reorganizing the exact quantization condition, 
we improve Gutzwiller's analysis in a crucial way by  bion contributions  (incorporating complex periodic paths) and turn it into an  exact result.     Furthermore, we argue the novel meaning of quasi-moduli integral and provide a relation between the Maslov index and the intersection number of Lefschetz thimbles.

\end{abstract}

\maketitle

\tableofcontents

\newpage

\section{Introduction}

In large variety of quantum mechanical systems, it is now well-understood  that  perturbative (P) and non-perturbative (NP) physics are  related in a deep way.   This connection can be understood by multiple means. These are,  
\begin{itemize}
\item Semi-classical analysis based on  saddle-point  (e.g. instantons) and steepest descent (Lefschetz thimble) methods. 
\item  Exact WKB method based on Schr\"{o}dinger  equation.
\item Exact quantization methods based on generalization of Bohr-Sommerfeld quantization.
\end{itemize}
In each one of these constructions,  to see the connection between P/NP  physics, the most prominent role is played by resurgence theory, and Stokes phenomena. 
Despite the fact that much is known about each one of these methods, the precise relation between them is not yet completely clear. 

In the resurgent asymptotic   analysis, see  \cite{Ec1,Pham1,BH1,Howls1,DH1,Costin1,Sauzin1,Sauzin2} for mathematical background, 
the   large order growth
of the perturbative coefficients of fluctuations about a given  sector (e.g. perturbative sector), is related to 
low-order perturbative coefficients of fluctuations about other non-perturbative sectors (e.g. the instanton-anti-instanton sector) in a precise way.    
Many signs  of resurgent structure in the perturbative and instanton analysis are already hinted  in the old physics literature 
\cite{ Bender:1969si, Bender:1973rz, 
Brezin:1977ab, Lipatov:1977cd, Bogomolny:1980ur, ZinnJustin:1981dx, ZinnJustin:1982td, ZinnJustin:1983nr,Aoyama:1991ca,Aoyama:1994sk,Aoyama:1995ca, Aoyama:1997qk,Aoyama:1998nt, ZinnJustin:2004ib, ZinnJustin:2004cg, Jentschura:2010zza, Jentschura:2011zza, Jentschura:2004jg}.  
The renewed interest is due to the precise understanding of the  connection between  resurgence theory and physical problems in quantum mechanics    \cite{Dunne:2013ada, Basar:2013eka,Dunne:2014bca,Escobar-Ruiz:2015nsa,Escobar-Ruiz:2015nsa2,Misumi:2015dua,Behtash:2015zha,Behtash:2015loa,Gahramanov:2015yxk,Dunne:2016qix,Fujimori:2016ljw,    
Sulejmanpasic:2016fwr,Dunne:2016jsr,Kozcaz:2016wvy,Serone:2016qog,Basar:2017hpr,Fujimori:2017oab,Serone:2017nmd,Behtash:2017rqj,Costin:2017ziv,Alvarez:2017sza,Fujimori:2017osz,Sueishi,Ito:2018eon,  Behtash:2018voa, Pazarbasi:2019web},  matrix models and string theory  \cite{Marino:2006hs,Marino:2007te, Marino:2008ya, Marino:2008vx, Pasquetti:2009jg,Garoufalidis:2010ya, Drukker:2010nc, Aniceto:2011nu, Marino:2012zq,Schiappa:2013opa,Hatsuda:2013oxa,Aniceto:2013fka,Santamaria:2013rua,Grassi:2014cla,Couso-Santamaria:2014iia,Grassi:2014uua,Couso-Santamaria:2015wga, Aniceto:2015rua,Dorigoni:2015dha,Hatsuda:2015qzx,Franco:2015rnr,Couso-Santamaria:2016vcc,Kuroki:2016ucm,Couso-Santamaria:2016vwq,Arutyunov:2016etw}, 
and  quantum field theory  \cite{Dunne:2012ae, Dunne:2012zk,Cherman:2013yfa,Cherman:2014ofa,Misumi:2014jua,Misumi:2014bsa,Misumi:2014rsa,Nitta:2014vpa,Nitta:2015tua,Behtash:2015kna,Dunne:2015ywa,Misumi:2016fno,Demulder:2016mja,Sulejmanpasic:2016llc,Fujimori:2018kqp,Okuyama:2018clk,Ishikawa:2019tnw,Yamazaki:2019arj,Ishikawa:2020eht,Morikawa:2020agf, Argyres:2012vv, Argyres:2012ka,Dunne:2015eoa,Dunne:2015eaa,Buividovich:2015oju,Dunne:2016nmc, Gukov:2016njj, Yamazaki:2017ulc,Ashie:2019cmy,Ishikawa:2019oga,Aniceto:2014hoa,Honda:2016mvg,Honda:2016vmv,Dorigoni:2017smz,Fujimori:2018nvz}.\footnote{In this paragraph, we are referring to the more standard  version of resurgence, which is a network of  ``large-order/low order"  relations.  A   constructive version of resurgence, based on  ``low-order/low order"  relations is discussed in \cite{Alvarez1,Alvarez2,Alvarez3,  Dunne:2013ada, Dunne:2014bca}.}   The power of formalism comes from the fact that the resurgent structure can provide a complete non-perturbative definition of quantum theories, reveal new saddles that are not obvious at all, and may provide at least a partial  solution to the important  renormalon problem in QFTs \cite{Dunne:2012ae, Argyres:2012vv}

In the semi-classical saddle point approach, we express physical quantities such as ground state energy or the partition function as series containing perturbative and non-perturbative contributions,  \footnote{Instead of partition function, one may also consider generalized partition function with the insertion of an operator leading to transition between different vacua, such as $\tr (e^{-\beta H} P)$ in parity invariant systems. 
In such cases, the transseries will begin with an instanton factor instead of perturbative vacuum.}
\begin{align}
	Z(\hbar)=  \sum_n a_n\hbar^n+e^{-\frac{S_1}{\hbar}}\sum_nb_n\hbar^n+e^{-\frac{S_2}{\hbar}}\sum_nc_n\hbar^n+...
\end{align}
This type of series is called as a  trans-series.
The perturbative and non-perturbative parts are usually calculated independently, and typically, all series appearing in the transseries are asymptotic expansions.  
One can turn the divergent  asymptotic series into something finite by Borel resummation, but the 
price one pays is that the result may be   multi-fold ambiguous. 
  For example, the Borel resummation of the perturbative part  
$ \sum a_n\hbar^n$  induces an ambiguity  which has a   non-perturbative factor $e^{-{S_1}/{\hbar}}$ (where $S_{1}$ is   an information about another saddle in the problem,  associated  with  the   instanton antiinstanton action) is expressed as
\begin{align}
	(\mathcal{S}_+ -\mathcal{S_-})\left[\sum_{n}a_n\hbar^n\right]\,\propto\,  ie^{- \frac{S_1}{\hbar}}\,,
\end{align}
Here ${\mathcal S}_{\pm}$ stands for the operation of lateral  (left/right) Borel summation, in which the sign $\pm$ means how the Laplace integral contour in the Borel resummation avoids the Borel singularity.  
This phenomenon is considered to be equivalent to the Stokes phenomena in the Picard-Lefschetz theory, where the structure of the Lefschetz thimble decomposition  changes  discontinuously \cite{ Cherman:2014ofa, Witten:2010cx,Cristoforetti:2013wha,Fujii:2013sra,Tanizaki:2014tua,Tanizaki:2014xba,Kanazawa:2014qma,Tanizaki:2015tnk,DiRenzo:2015foa,Fukushima:2015qza,Tanizaki:2015rda,Fujii:2015bua,Alexandru:2016gsd,Tanizaki:2016xcu}. 
From a physical point of view, it means that the perturbative part, even after Borel resummation, is not well-defined by itself. But crucially,  the way it is not well-defined carries 
  non-perturbative information encoded into it.  For stable quantum mechanical systems,  the energy eigenvalues must be   real and unambiguous. Fortunately, there are also non-perturbative contributions, instanton-anti-instanton correlated events $e^{-\frac{S_1}{\hbar}}\sum b_n\hbar^n$, whose contributions are  also two-fold ambiguous and cancels the ambiguity of Borel resummed perturbative series. This is first calculated 
  in   Refs.~\cite{Bogomolny:1980ur, ZinnJustin:1981dx}, but rigorous explanation is given in \cite{Behtash:2018voa} by using the concept of critical points at infinity and   Picard-Lefschetz theory. 
This type of resurgent  cancellations 
 encodes an intricate network of  ``large-order/low order" relations between different perturbative and non-perturbative sectors, and is  partially proven.

In the  exact-WKB analysis \cite{Balian:1978ab, Voros1, DDP1, CNP1, DLS1, DP1, Takei1, CDK1, Takei2, Getm1, AKT1, Schafke1, Getm2,  Iwaki1, Hollands:2019wbr, Kashani-Poor:2015pca, Ashok:2016yxz}, 
which has been studied mainly by mathematicians,  one investigates the properties of solutions to certain differential equations using the Borel summation.\footnote{Of course, the usual  WKB approximation is a text-book material. But it turns out that standard WKB can be made exact as we review here.  
The exact-WKB, for which resurgent analysis is the fundamental tool, is more recent   (a clarified understanding began with the works of Voros and Silverstone in the 80's \cite{Silverstone, Voros1})  and there is a very interesting  body of mathematical works around it. One of our goals here is to make it more accessible, and use it as a tool to explore the connections between  various non-perturbative methods.}
Our main interest is in its application to the Schr\"{o}dinger equation,
 \begin{align}
	\qty(-\frac{\hbar^2}{2}\frac{d^2}{dx^2}+V(x))\psi(x)=E\psi(x)\,. 
\end{align}
Here $\psi(x)$ correspond to the wave function and $E$ stands for the energy eigenvalue. In exact-WKB, one maps the Schr\"odinger equation into a non-linear Ricatti equation, whose asymptotic solution provide the building block of  WKB-wave function, which is   by itself 
 an asymptotic series. In exact-WKB, the position   is first elevated to a complex variable, and the classical potential is used to  turn the 
 $x\in \mathbb C$ plane into  Stokes graph and regions.\footnote{In semi-classical approach to path integral formulation, the action and space of paths  must be complexified  from the beginning, and all (real and complex) saddles must be determined, see for example  \cite{Balian:1978ab, Grassi:2014cla} and \cite{Behtash:2015zha,Behtash:2015loa}. 
 This seems to be the counterpart of promoting   $x\in \mathbb R$ to 
  $x\in \mathbb C$ in exact WKB and determining all real and complex turning points. Whether a complex saddle contributes to observables or not 
  is not always easy to determine in path integral formulation, but this can be easily determined from the Stokes graph in exact WKB.} 
    Once one considers 
the  analytic continuation of the WKB-solution of $\psi(x)$ from one Stokes region to an adjacent  one  in complex  $x$  plane, 
its asymptotic behavior sometimes changes discontinuously.
\begin{align}
	\psi^+_{\textrm{I}}(x)\rightarrow\psi^+_{\textrm{I\hspace{-.1em}I}}(x)+\psi^-_{\textrm{I\hspace{-.1em}I}}(x)
\end{align}
This is another type of the Stokes phenomenon, and this change can be examined by studying the Borel summation of the WKB  wave functions. 
Starting with a decaying WKB wave function in the $x \rightarrow -\infty$, 
the existence of  Stokes jumps and connection formula induce both exponentially decaying {\it and} exponentially increasing components as 
 $x \rightarrow +\infty$. But  the WKB wave function must vanish as $x \rightarrow + \infty$  for normalizability.  This simple fact implies that 
 the prefactor of the exponentially increasing WKB-wave function must vanish. This is the  statement of   the  exact quantization condition.  

Exact quantization conditions necessarily involves both perturbative and non-perturbative cycles,  and forces  precise relations between perturbative and non-perturbative contributions. 
 We investigate implications of this construction, 
and interpret it in terms of Euclidean path integral formulation. This reveals 
how  the  two different kinds of Stokes phenomena in the two  ``seemingly" different formulations  are related. 
%


Let us briefly  summarize our findings:

\begin{enumerate}

\item
{\bf Unified understanding of the two Stokes phenomena:}
 As described above,  in the semiclassical analysis of the double-well quantum system, the Stokes phenomenon occurs in  such a way that 
  the imaginary ambiguities are canceled out between the perturbative Borel resummation and the non-perturbative bion contributions.  
  In the exact-WKB analysis, we find the same Stokes phenomenon takes place as the change of the ``topology'' of the Stokes curve. This correspondence is clearly incorporated in the Delabaere-Dillinger-Pham (DDP) formula \cite{DDP1}, which expresses the resurgent relation as a relation between the different cycles crossing the Stokes curves.

\item 
{\bf Generalizing the Gutzwiller trace formula:} 
The Gutzwiller trace formula is a  semi-classical method that ties  spectrum  of quantum theory to classical mechanical concepts, to periodic orbits calculations, actions, geometric phases  \cite{Gutzwiller}. However, there was no unified way to determine which periodic solution should be added up as a unit (prime periodic orbit)\footnote{In 2018, Nekrasov \cite{Nekrasov:2018pqq} suggested that 
Gutzwiller's formula can be improved by the contributions of what he calls {\bf  m,n}-solutions (concrete examples of which are bion configurations as he points out), such that it can produce an exact formula. Our findings proves this  proposal at least for a number of polynomial potentials.  
} and how to sum up the units.  In this paper, we discover the uniform way to identify the unit orbits and how to sum them up with including instanton effects.

\item
{\bf Novel meaning of quasi-moduli integral:}  The above findings give new physical meaning to the quasi-moduli integral (QMI) in the semiclassical analysis of path integral. Using the perspective of Gutzwiller's quantization condition, the non-perturbative contribution obtained from QMI is shown to have a nontrivial relation with the perturbative contribution around the classical vacuum.

\item {\bf Discovering the relation between Maslov index and the intersection number of Lefschetz thimble:}
The resolvent $G(E)$ obtained from the quantization condition $D(E)=0$ in the exact-WKB analysis can be rewritten in the Gutzwiller-type representation and it can be compared to $G(E)$ derived in the Gutzwiller's quantization condition. We then find that $(-1)^n$ appearing in the non-perturbative sector is interpreted as the Maslov index.
Furthermore, we show the Maslov index turns out to be the intersection number of Lefschetz thimbles by expressing   the Fredholm determinant and resolvent in a convenient form.

\item {\bf Equivalence of the different quantization conditions:}
Based on above observations, we clarify the equivalence and the nontrivial relations among the different quantization conditions, based on path integral, exact-WKB and Gutzwiller methods.

\item {\bf Generalization to symmetric multi-well potential:}
We generalize the exact-WKB analysis and the DDP formula to the quantum systems with generic symmetric multi-well potentials. Again, we show that the Stokes phenomena occur as the topological change of Stokes curve in the exact-WKB analysis, and the resurgent structure in the semiclassical analysis is completely incorporated in the DDP formula.

\end{enumerate}

The paper is constructed as follows.
In Sec.~\ref{sec:PI} we review path integral, Lefschetz thimble decompositions, resolvent methods and Gutzwiller's quantization in quantum theories.
In Sec.~\ref{sec:EWKB} and Sec.~\ref{sec:HO} we review the exact-WKB method, and explain Stokes curves for potential problems in simple examples. 
In Sec.~\ref{sec:DW} we apply the exact-WKB analysis to double-well potential quantum mechanics by studying the associated Stoke curves,
where we find the equivalence of the two Stokes phenomena and show the equivalence among several quantization conditions.
In Sec.~\ref{sec:Generic} we extend our investigation to the systems with generic multi-well potentials and discuss outcomes.
Sec.~\ref{sec:summary} is devoted to summary and discussion.

\section{Preparation}
Before starting our journey, we introduce the tools other than the exact-WKB analysis as prerequisite knowledge. These include  saddle point decomposition of path integrals (Lefschetz thimbles), its relation to resolvent, 
 Gutzwiller's quantization and Maslov index. Our discussion is basic and  is streamlined according to what we need later.

\subsection{Lefschetz thimble decomposition and Resolvent method}
\label{sec:PI}

We start with the Lefschetz-thimble decomposition of path integral and the resolvent method in quantum mechanics.
By use of the asymptotic series expansion and the trans-series expansion, the Euclidean partition function in quantum mechanical systems with field   $x(\tau)$ is expressed as
\begin{align}
Z(\beta)  =\int \mathcal{D}x\;e^{-\frac{S[x]}{\hbar}}
 = \sum_n a_n\hbar^n + e^{-\frac{S_1}{\hbar}}\sum_n b_n\hbar^n + e^{-\frac{S_2}{\hbar}}\sum_n c_n\hbar^n+... \,,
\end{align}
where  $\beta$ is Euclidean time period. 
From the viewpoint of Picard-Lefschetz theory,  the    Borel summation of a 
perturbative series around each saddle point   corresponds to  performing  exact thimble integration 
associated with the corresponding saddle:\footnote{More precisely, the perturbative series itself (without the Borel summation)  corresponds to expanding the interaction terms in action into a series  and performing Gaussian integration over the quadratic field. Naturally, this procedure gives an asymptotic divergent expansion. The Borel sum of this series is exactly equal to the integration over thimble. 
If the thimble decomposition is unambiguous, the Borel sum is also unambiguous, and series is Borel summable. A  Borel ambiguity happens when there is a Stokes phenomenon, i.e,  a topology change in the thimbles and their  decomposition. The ambiguity in the thimble decomposition is the same ambiguity in the Borel resummation.}

\begin{align}
Z(\beta) \,=\, \sum_\sigma n_\sigma \int_\mathcal{J_\sigma} \mathcal{D}x\; e^{-\frac{S[x]}{\hbar}}\,, 
\end{align} 
Here $\sigma$ labels the saddle points and $\mathcal{J_\sigma}$ are the corresponding Lefschetz thimbles. 
Here, we have to determine  the index, $n_\sigma$($0$ or $\pm 1$), called the intersection number, to obtain the correct result of the path integral. However, there is no efficient method for calculating this index except for calculating the thimble numerically and plotting it explicitly, which is a hard task.  Therefore, we have no reliable way to determine ``which are relevant saddles" in the generic cases from the integration itself. In this paper, we propose a certain solution to this problem by using exact-WKB and simple Stokes graphs,   and give a physical interpretation of the index in quantum mechanics.

We now review the resolvent method~\cite{ZinnJustin:2004ib} for quantum theories. In the latter part of this paper, this method will enable us to obtain the partition function of the system directly from the quantization conditions obtained from the exact-WKB analysis and to interpret the resurgent structure of the partition function in terms of the exact-WKB analysis.

First, we write down the partition function  formally as a sum over saddle points
\begin{align}
Z(\beta)  &=\tr e^{-\beta\hat{H}}  =\int \mathcal{D}x\;e^{-\frac{S[x]}{\hbar}}                                                         \nl
& = n_0  \; \mathcal{S}\left[e^{-\frac{S[x_0]}{\hbar}}\sum_n a_n\hbar^n \right]+ n_1 \; \mathcal{S}\left[e^{-\frac{S[x_1]}{\hbar}}\sum_n b_n\hbar^n \right]+...                  \nl
& =\sum_\sigma n_\sigma \; \int_{\mathcal{J}_\sigma} \mathcal{D}x\;e^{-\frac{S[x]}{\hbar}}=\sum_\sigma n_\sigma\;Z_\sigma(\beta)\,, 
\end{align}
where $\mathcal{S}[\cdot]$ denotes the Borel summation of series expansions and $x_\sigma$ stands for saddle points.  

We then consider the Laplace transform of $Z(\beta)$, which gives the trace of resolvent $G(E)$.
Since this transform is linear, we obtain the expression as 
\begin{align}
\tr\frac{1}{\hat{H}-E}=G(E) & =\int_0^\infty Z(\beta)e^{\beta E}\dd{\beta}                             \nl
	                      & =\sum_\sigma n_\sigma \int_0^\infty Z_\sigma(\beta)e^{\beta E}\dd{\beta} \nl
	                      & =\sum_\sigma n_\sigma G_\sigma(E)\,.                                     
	\label{eachresolvent}
\end{align}
It is notable that the poles of $G(E)$ give the eigenvalues and $G_\sigma(E)$ stands for the trace of resolvent for each sector (each thimble).
The trace of resolvent $G(E)$ can be connected to the Fredholm determinant $D(E)=\det(\hat{H}-E)$ 
via the relation  $-\pdv{E}\log D=G(E)$. 
Then, we have
\begin{align}
	D(E)=\prod_\sigma D_\sigma^{n_\sigma}(E)\,,
\end{align}
where $D_\sigma(E)$ stands for the Fredholm determinant for each  thimble.
We note that the zeros of $D(E)$ give the exact energy eigenvalues.\footnote{Mathematically, the definition of Fredholm determinant (or resolvent) needs a regularization, e.g. $G_{reg.}\equiv G(E)-G(0)$ or $D_{reg.}\equiv \frac{D(E)}{D(0)}$ or zeta function regularization for $D(E)$. }
 

The main point of  our analysis is following. From the exact-WKB analysis, we will obtain an exact  quantization condition, $D(E)=0$.  This formula will be expressed in terms of perturbative and non-perturbative cycles, which involve perturbative as well as non-perturbative instanton/bion data. 
By reexpressing the condition $D(E)=0$ as a sum over P/NP cycles,  we will be able  
extract  the index $n_\sigma$ and Maslov index from the exact-WKB analysis. 
%
Since we can go back to $Z(\beta)$ by inverse Laplace transform
\begin{align}
	G(E)     & =\int_0^\infty Z(\beta)e^{\beta E} d\beta                                          \\
	Z(\beta) & =\frac{1}{2\pi i}\int_{\epsilon-i\infty}^{\epsilon+i\infty}G(E)e^{-\beta E} dE\,,
\end{align}
it will enable us to obtain the partition function from the exact-WKB method.

\subsection{Gutzwiller's quantization}
\label{sec:Gutzwiller}
Gutzwiller's quantization, which is also known as \textit{the Gutzwiller trace formula}\cite{Gutzwiller}, is a semi-classical construction that express the quantum mechanical density of states (the resolvent: $G(E)$), in terms of periodic orbits. The formalism  uses  path integral in Minkowski space formulation (in real time), hence one is dealing with amplitudes in real time $\mathcal{D}x\;e^{iS}$.  
In certain sense, Gutzwiller method can be interpreted  as the intermediate quantization method between the path integral and the Bohr-Sommerfeld quantizations. Actually, the distribution of the poles of $G(E)$, which determines the energy eigenvalues, gives the Bohr-Sommerfeld quantization condition. Later, we will show how to derive this trace formula interms of exact-WKB method and resurgence.   

We first express the ``Lorentzian" partition function as
\begin{align}
Z(T) =\tr e^{-i\hat{H}T}                        
=\int_{periodic}\mathcal{D}x\;e^{iS} \,,
\end{align}
The resolvent, which has quantum spectral data,  is given by 
\begin{align}
G(E)  =\int_0^\infty Z(t) e^{(iE-\epsilon)T} dT   =\int_0^\infty \sum_n e^{(iE-iE_n-\epsilon)T}dT  =-i\tr\frac{1}{\hat{H}-E}\,.                      
\end{align} 
where $\lim_{\epsilon \searrow 0} $ is taken after integration. 
The resolvent can  also be  expressed as
\begin{align}
G(E)=-i\tr\frac{1}{\hat{H}-E}  =\int_0^\infty dT \int_{periodic}\mathcal{D}x\;e^{iS+iET}  
=\int_0^\infty dT \int_{periodic}\mathcal{D}x\;e^{i\Gamma}\,, 
\label{GutzwillerGamma}
\end{align}
where $\Gamma=S+ET$. We also note that the action $S$ is written as
\begin{align}
S  =\int^T p \dot{x} dt-\int^T H dt  =\oint p dx -\int^T H dt\,.
\end{align}
We here evaluate the $T$ integral in~(\ref{GutzwillerGamma}) by the stationary phase method with considering the $T$ derivative of $\Gamma$
\begin{align}
\dv{\Gamma}{T}=\dv{S}{T}+E\,,
\end{align}
Since $\oint pdx$ is the area of phase space, which depends on the trajectory but not on $T$ (how long it takes to go around), i.e. $\dv{T}\oint pdx=0$.
Therefore, we obtain
\begin{align}
\dv{\Gamma}{T}=\dv{S}{T}+E=-H+E \,.
\end{align}
It means the leading contributions of the $T$ integral are periodic classical solutions whose energy is $E$. When a periodic orbit is a solution, the configuration rotating $n$ times is also a solution. By taking this fact into account, we find that the contribution is obtained just by the replacement as $\oint pdx\rightarrow n\oint pdx$. 
Then, we obtain
\begin{align}
\Gamma=S+ET=\qty(n\oint pdx-ET)+ET=n\oint pdx\,,\quad\;\;\;(n=1,2,3...) \,.
 \end{align}
After all, the contribution of the classical solutions to $G(E)$ is expressed as
\begin{align}
G(E)\simeq\sum_{p.p.o.} \sum_{n=1}^\infty e^{in\oint_{p.p.o.} pdx}\,, 
\end{align}
where $p.p.o.$ stands for a ``prime periodic orbit", which is a topologically distinguishable orbit among the countless periodic orbits.

If we consider the sub-leading terms in  stationary phase approximation, it gives
\begin{align}
G(E)\simeq\sum_{p.p.o.}\sum_{n=1}^\infty \exp\left(i\,n\oint_{p.p.o.} pdx\right)\,\qty(\det\frac{\delta^2S}{\delta x\delta x})^{-1/2} \,,
\end{align}
Here, $\det\frac{\delta^2S}{\delta x\delta x}$ is a functional determinant taking into account the fluctuation operator around the saddle point.  Evaluation of this part requires care, 
as described below, this operator has negative eigenvalues when considering the expansion around a periodic orbit in general. The number of negative eigenvalues is called ``Maslov index", 
which plays an important role in Gutzwiller's quantization.

\subsubsection{Maslov index}
  \label{sec:AppendixMaslov}
  Let $x_{cl}$ denote the classical solution and $\delta x $ denote the fluctuations around it. The integration over  fluctuations 
at the quadratic level is determined by the functional determinant of the fluctuation  operator:
  \begin{align}
    M=\frac{\delta^2S}{\delta x\delta x}=-\dv[2]{t}-V''(x_{cl})\,.
  \end{align}
The operator $M$ has a zero eigenvalue if $x_{cl}$ depends on $t$, and the operator $M$ has 2n-1 negative eigenvalues for n-cycle orbit. 
We below give a brief proof of this fact:

\begin{proof}
    Consider classical EoM:
    \begin{align}
      -\dv[2]{x_{cl}}{t}-\dv{V}{x_{cl}}= 0\,.
    \end{align}
    Take $t$ differential for this equation. Then we get
    \begin{align}
      \qty(-\dv[2]{t}-V''(x_{cl}))\dv{x_{cl}}{t} = 0\,.
    \end{align}
This expression is nothing but an eigenvalue equation for the zero eigenvalue of the fluctuation operator,  $M \tilde \psi_0(t)=0$, and the eigenfunction is proportional to $ \tilde \psi_0(t)= \dv{x_{cl}}{t}$.

Next, let us consider a periodic classical solution $x_{cl}$. When it is a one-cycle solution, the derivative $\dv{x_{cl}}{t}$ typically has a behavior depicted in Fig.~\ref{fig:dxcl}.
\begin{figure}
\centering
\includegraphics[width=5cm]{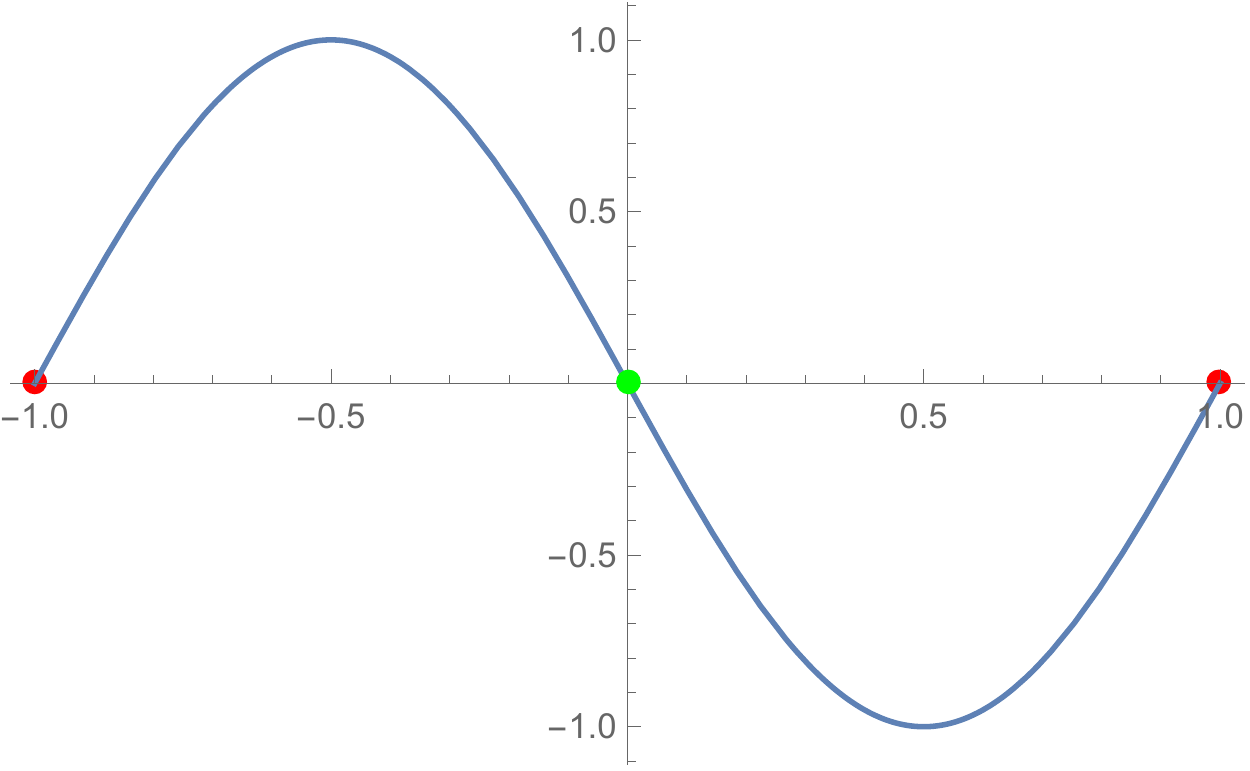}
\caption{The appearance of the derivative $\dv{x_{cl}}{t}$ for 1-cycle. }
\label{fig:dxcl}
\end{figure}
The operator $M$ is a Schr\"{o}dinger-type operator, thus the level of the eigenfunction is determined by the number of zero points\footnote{If the eigenfunction of $M$ is on $\mathbb{R}^1=(-\infty,\infty)$, the level is same to the number of nodes but when $S^1$, the level is the number of nodes -1.} (nodes). In the case of Fig.~\ref{fig:dxcl}, $M$ has two nodes since the periodic b.c. is imposed, and the endpoints are identical and regarded as a single node.
The reason why it has one negative eigenvalues is that $\dv{x_{cl}}{t}$ is the first excited state, but at the same time, $\dv{x_{cl}}{t}$ is also an eigenfunction of the zero eigenvalue.
Similarly, a $n$-cycle classical solution has $2n$ turning points, $M$ has $2n-1$ negative eigenvalues because $\dv{x_{cl}}{t}$ is the $2n-1$-th excited state.
\end{proof}

If we consider an analogy with the Morse theory, the operator $M$ corresponds to (the diagonal part of) Hessian where  the action is viewed as a Morse function, and its negative eigenvalue corresponds to the Morse index.
Thus, the Maslov index is essentially regarded as the Morse index in the functional integral.
To rephrase this, we now express the contribution of functional determinants as
\begin{align}
\sqrt{\det M}=\sqrt{|\det M|}e^{i\alpha\pi},\quad\quad\quad\alpha=\frac{\nu}{2}\,.
\end{align}
Here,  $\alpha$ is called the  \textit{Maslov index}. Here, $\nu$ is the number of negative eigenvalues of $M$. 
The determinant of the $n$-cycle is given by
\begin{align}
\sqrt{\det M}=-i\sqrt{|\det M|}(-1)^n\,.
\end{align} 
Therefore, the final form of $G(E)$ (up to higher order quantum corrections) is given as
\begin{align}
G(E)=i\sum_{p.p.o.}\sum_{n=1}^\infty T(E)\,e^{in\oint_{p.p.o.} pdx}(-1)^n\qty( \qty| \det \frac{\delta^2S}{\delta x\delta x}|)^{-1/2}\,,
\label{Gutzwiller_form} 
\end{align}
where $T(E)$ is the period of each cycle, which comes from the zero eigenvalue of $M=\frac{\delta^2S}{\delta x\delta x}$. Also, we call $(-1)^n$ as Maslov index instead of $\alpha$ in the latter calculation.

\noindent
{\bf Working of Maslov index in simple harmonic oscillator:}
As an example, we now consider the harmonic oscillator system.
In this case, there is only one type of p.p.o. with constant $T(E)$ and $|\det \frac{\delta^2 S}{\delta x_i\delta x_j}|$,
We then obtain
\begin{align}
G(E)  \propto\sum_{n=1}^\infty e^{in\oint pdx}(-1)^n =\frac{e^{i\oint pdx}}{1+e^{i\oint pdx}}\,.      
\end{align}
Therefore the poles of $G(E)$ are given by
\begin{align}
\oint pdx=2\pi\qty(n+\frac{1}{2})\,.
\end{align}
This is the Bohr-Sommerfeld quantization of harmonic oscillator. 
It should be emphasized that the contribution of the Maslov index is important 
in order to obtain the correct energy eigenvalues including the vacuum energy.

However, the way to determine p.p.o. in this method is not well understood in the most of cases and almost exclusively used in systems without tunneling phenomena (instantons) \cite{Gutzwiller}.  
In retrospect, this is not surprising because usual instantons in real time correspond to imaginary singular configurations \cite{Cherman:2014sba}, and it is not so obvious how to deal with it. 
As we will show later, our finding gives the systematic method to determine p.p.o. including instanton-like configurations without any approximation. Furthermore, it also shows the relation between Maslov index and the intersection number of Lefschetz thimble.

%
%

\vspace{-0.6cm}
\section{Exact WKB}
\label{sec:EWKB}
In this section, we review the exact WKB method \cite{Balian:1978ab,Voros1,DDP1,CNP1,DLS1,DP1,Takei1,CDK1,Takei2,Getm1,AKT1,Schafke1,Getm2,Iwaki1, Hollands:2019wbr,Ito:2018eon} and the related techniques, including Borel resummation, Stokes curves and monodromy matrices. \footnote{Although in standard quantum mechanics books 
WKB is presented as an approximation which applies to high-energy states, this perspective is not correct. It is an exact method, 
and applies every where in the spectrum. }
For simplicity, we focus on the one-dimensional Schr\"{o}dinger equation,  and assume that  the potential $V(x)$ doesn't include $\hbar$, i.e. it is a purely classical potential. 
\begin{align}
	\qty(-\frac{\hbar^2}{2}\frac{d^2}{dx^2}+V(x))\psi(x)=E\psi(x)\,. 
\end{align}
We set $Q(x)=2(V(x)-E)$ then rewrite the equation as
\begin{align}
	\label{SchroedingerQ}                              
	\qty(-\frac{d^2}{dx^2}+\hbar^{-2}Q(x))\psi(x)=0\,, 
\end{align}
In the WKB analysis, we consider the ansatz given by
\begin{align}
	\psi(x,\hbar) & =e^{\int^x S(x,\hbar)dx} \,,                                     
	\label{WKBpsi}
	\\
	S(x,\hbar)    & =\hbar^{-1} S_{-1}(x)+S_0(x)+\hbar S_1(x)+\hbar^2 S_2(x)+...\,, 
	\label{WKBS}
\end{align}
where  $	S(x,\hbar)  $ is a formal power series expansion in  expansion parameter $\hbar$, and  $S_n(x)$ are functions of $x$.  
Substituting  Eq.~(\ref{WKBpsi}) into Eq.~(\ref{SchroedingerQ}), 
leads  to the non-linear  Riccati equation
\begin{align}
	S(x)^2+\pdv{S}{x}=\hbar^{-2}Q(x)\,. 
	\label{Riccati}                     
\end{align}
By substituting Eq.~(\ref{WKBS}) into Eq.~(\ref{Riccati}), we obtain the recursive relation
\begin{align}
	\label{RiccatiRecursive}                                                   
	S_{-1}^2=Q(x)\,,                                                           
	\quad\quad\quad                                                            
	2S_{-1}S_n+\sum_{j=0}^{n-1}S_jS_{n-j}+\pdv{S_{n-1}}{x}=0\;\;\;(n\geq 0)\,. 
\end{align}
We note $S_{-1}=\pm \sqrt{Q(x)}$. Since $S_n$ is recursively determined from $S_{-1}$, $S_n$ has two independent solutions:
\begin{align}
	S^{\pm}(x,\hbar)=\hbar^{-1} S^{\pm}_{-1}(x)+S^{\pm}_0(x)+\hbar S^{\pm}_1(x)+\hbar^2 S^{\pm}_2(x)+...\,, 
\end{align}
The first several terms are given by 
\begin{align}
	S^{\pm}_{-1}(x) & =\pm\sqrt{Q(x)}\,,                                                                          \\
	S^{\pm}_{0}(x)  & =-\frac{\pdv{Q}{x}}{4Q}\,,                                                                  \\
	S^{\pm}_{1}(x)  & =\pm\qty(-\frac{5}{32}\frac{\qty(\pdv{Q}{x})^2}{Q^{5/2}}+\frac{\pdv[2]{Q}{x}}{8Q^{3/2}})\,. 
\end{align}
From Eq.~(\ref{RiccatiRecursive}), one finds the relation $S_n^{-}=(-1)^nS_n^+$. 
Therefore, we reach the simple expression
\begin{align}
	S^\pm(x,\hbar) & =\hbar^{-1} S^{\pm}_{-1}(x)+S^{\pm}_0(x)+\hbar S^{\pm}_1(x)+\hbar^{2}S^{\pm}_2(x)+... \\
	               & =\pm\hbar S_{-1}^+ +S_0^+ \pm \hbar S_1^+ +\hbar^2 S_2^+ +...                         \\
	               & =\pm S_{{\rm odd}}+S_{{\rm even}}\,.                                                              
	\label{Soddeven}
\end{align} 
Based on Eq.~(\ref{Soddeven}), Eq.~(\ref{Riccati}) is rewritten as
\begin{align}
	(S_{{\rm odd}}+S_{{\rm even}})^2+\pdv{x}\qty(S_{{\rm odd}}+S_{{\rm even}})=\hbar^{-2} Q\,,   \\
	(-S_{{\rm odd}}+S_{{\rm even}})^2+\pdv{x}\qty(-S_{{\rm odd}}+S_{{\rm even}})=\hbar^{-2} Q\,. 
\end{align}
These two equations give
\begin{align}
	\therefore S_{{\rm even}}=-\frac{1}{2}\pdv{x}\log S_{{\rm odd}} \,. 
\end{align}
Therefore the WKB wave function can be  expressed as
\begin{align}
	\psi^\pm_a(x)=e^{\int^x S^\pm dx} =\frac{1}{\sqrt{S_{{\rm odd}}}}e^{\pm\int^x_{a} S_{{\rm odd}} dx}\,, 
\end{align}
with $a$ being an integral constant.
For later calculations, we choose it as a turning point, which is a solution of $Q(x)=0$.
At the leading order, WKB wave function  is evaluated as
\begin{align}
	\psi^\pm_a(x)=\frac{1}{Q(x)^{1/4}}e^{\pm\frac{1}{\hbar}\int_{a}^x}\sqrt{Q(x)} dx\,, 
\end{align}
which is nothing but the solution in the text-book level  WKB approximation.

Since we have derived the WKB wave function recursively, it is regarded as a formal series in  $\hbar$ 
\begin{align}
	\psi^\pm_a(x) & =e^{\pm\frac{1}{\hbar}\int_{a}^x \sqrt{Q(x)}dx}\sum_{n=0}^{\infty}\psi_{a,n}^\pm(x)\hbar^{n+\frac{1}{2}}\,, 
	\\
	S_{{\rm odd}}       & =\sum_{n=0}^\infty S_{2n-1}\hbar^{2n-1}\,.                                                                  
\end{align}
Note that the factor $\frac{1}{2}$ in $\hbar^{n+\frac{1}{2}}$ comes from $\frac{1}{\sqrt{S_{{\rm odd}}}}$.
Here, both of these series  turn out to be asymptotic expansions with respect to $\hbar$.  In other words, 
the all orders  WKB wave function is an divergent asymptotic expansion with respect to $\hbar$. In order to give it a precise meaning, 
 we need another technology,  the Borel resummation,   applied to series for which the  divergent coefficients 
$ \psi_{a,n}^\pm(x)$ are $x$-dependent.

\subsection{Borel summation}
\noindent
Let us consider the following formal  series  (not necessarily asymptotic) with respect to $\hbar$.
\begin{align}
	Z(\hbar)=e^{-\frac{A}{\hbar}}\sum_{n=0}^{\infty}a_n\hbar^{n+\alpha}\;\;\;\;\alpha\notin \{-1,-2,-3,...\}\,. 
\end{align}
The Borel transform of this series is defined as
\begin{align}
	{\frak B}[Z](z) \equiv \sum_{n=0}^{\infty}\frac{a_n}{\Gamma(n+\alpha)}(z-A)^{n+\alpha-1}\,, 
\end{align}
The directional Borel resummation $\mathcal{S}[Z]$  is defined as
\begin{align}
	\mathcal{S}[Z](\hbar)\equiv\int_A^{\infty e^{i\theta}} e^{-\frac{z}{\hbar}}{\frak B}[Z](z)\dd{z}\;\;\;\;\theta={\rm Arg}(\hbar)\,. 
	\label{Borel2}
\end{align} 
where $\theta$ denotes the direction of integration. 

The resurgence property tells us that the Borel transform admits an analytic continuation in $z$ plane, which allows   via Borel resummation   the reconstruction of the exact value of the result. 
If the series is convergent, this  procedure just return the original series due to the identity
\begin{align}
	1=\frac{1}{\Gamma(n+\alpha)}\int_0^\infty e^{-x}x^{n+\alpha-1} dx\,. 
\end{align}
For an asymptotic  divergent series, however, it gives one  analytic functions which have
the series as its asymptotic series.

\begin{figure}[t]
	\centering 
	\includegraphics[width=13cm]{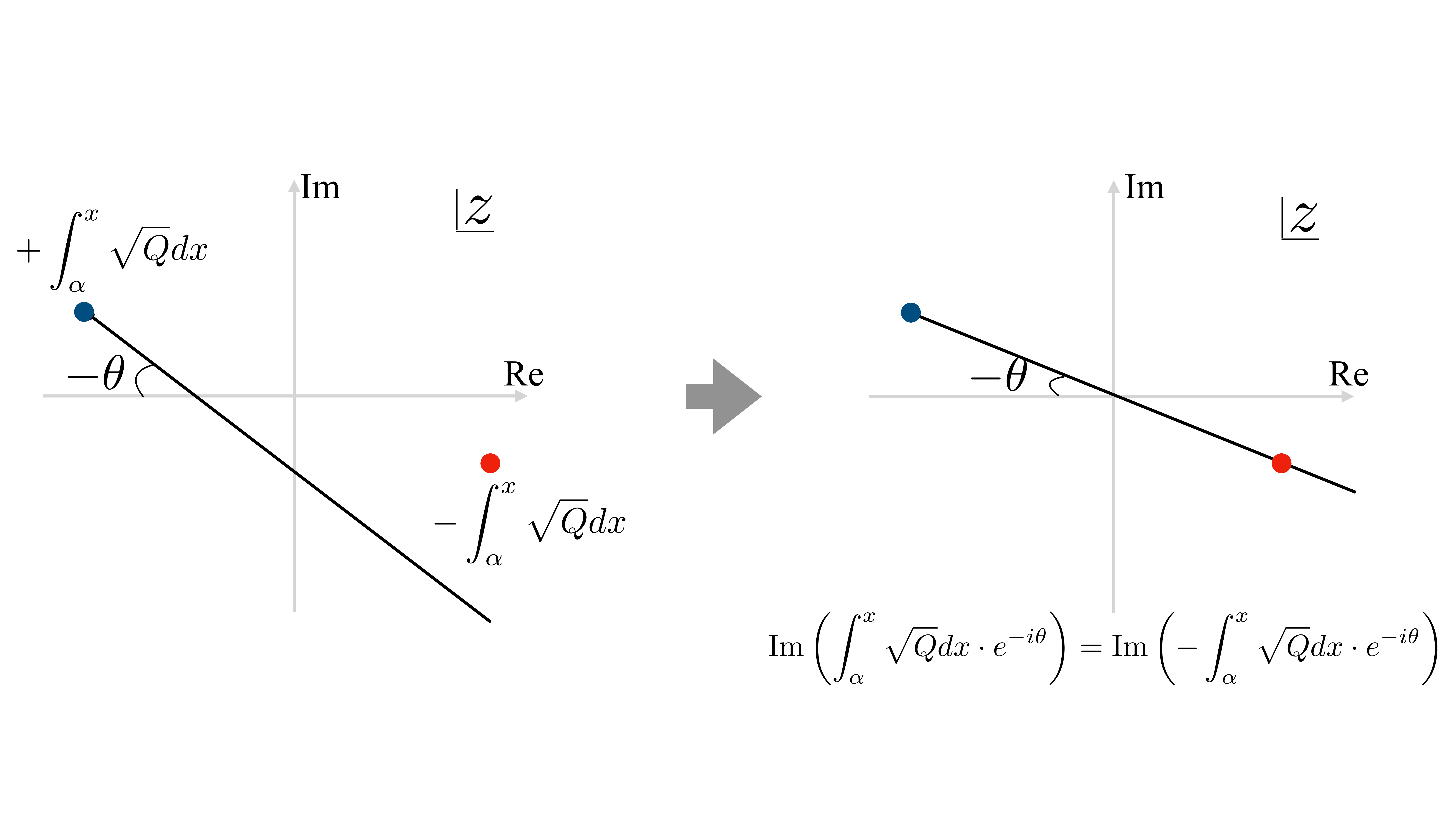}
	\caption{The black line indicates the integration path on the Borel plane, the blue circle is the endpoint of the integral path and the red circle indicates the singularity of ${\frak B}[\psi^\pm_a(x)](z)$.}
	\label{StokesCondition_picture}
\end{figure}

The Borel summation is a homomorphism, so that the following algebraic properties hold.
\begin{align}
	  & \mathcal{S}[A+B]=\mathcal{S}[A]+\mathcal{S}[B] \\
	  & \mathcal{S}[AB]=\mathcal{S}[A]\mathcal{S}[B]   
\end{align}
Now, we apply the Borel summation procedure to the WKB wave function, then we obtain
\begin{align}
	\mathcal{S}[\psi^\pm_a](\hbar) & =\int_{\mp z_0}^{\infty e^{i\theta}} e^{-\frac{z}{\hbar}} {\frak B}[\psi^\pm_a(x)](z)dz,\;\;\;\;\theta={\rm Arg}(\hbar)\,,                  \\
	{\frak B}[\psi^\pm_a(x)](z)    & =\sum_{n=0}^{\infty}\frac{\psi^\pm_{a,n}(x)}{\Gamma\qty(n+\frac{1}{2})}(z\pm z_0)^{n-\frac{1}{2}},\;\;\;\;z_0=\int_a^x\sqrt{Q(x)}\,. 
\end{align}
Because the coefficient $\psi^\pm_n(x)$ depends on $x$, the position of Borel singularity also depends on $x$. The position of singularities of integrand is $z=\pm\int_a^x\sqrt{Q(x)}$ when the Stokes curve is not degenerate. From now on, we will express the Borel-summed wave function $\mathcal{S}[\psi^\pm_a](\hbar)$ as just $\psi^\pm_a(\hbar)$ unless otherwise noted.

We now look into details of the Borel singularities.
One of them is always the endpoint of the integration  path, and $\int\frac{1}{\sqrt{x}}$ is regular at $x=0$. Therefore the other one contributes to Stokes phenomena. These two singularities are point-symmetric. 
Hence, Stokes phenomena occur due to the situations that one of the singularities is on the integration  path as shown in Fig.~\ref{StokesCondition_picture}.  This condition can be expressed as
\begin{align}
	\Im e^{-i\theta}\int_a^x\sqrt{Q(x)}dx=0 \,,              \nl
	\therefore \Im \frac{1}{\hbar}\int_a^x\sqrt{Q(x)}dx=0 \,.
	\label{StokesCondition}                               
\end{align}
Note that $\frac{1}{\hbar}\int_a^x\sqrt{Q(x)}=\frac{1}{\hbar} S_{-1}$ is the leading term of WKB expansion. Therefore, in order to understand when Stokes phenomena occur, we do not have to calculate Borel summation of $\psi(x)$ explicitly, but just evaluate this term.  The path derived from Eq.~(\ref{StokesCondition}) is called a  Stokes curve, which is part of a structure that is called Stokes graph.  In principle,     the exact  energy spectrum  of the theory can be calculated  just from the Stokes curve data.   We denote by ${\cal S}_{\pm}$   the lateral Borel resummation with a positive/negative (small) angle $\theta$.

\subsection{Stokes curves and Stokes phenomena}

Let $a$ be a turning point (a solution of $Q(x)=0$). In this case, the Stokes curve associated with $a$ is defined as
\begin{align}
	\Im \frac{1}{\hbar}\int_a^x \sqrt{Q(x)} dx=0 \,.
\end{align}
Also,  each  segment of the  Stokes curve has an index, $\pm$.  This index indicates which one of the  $\psi^+$ and $\psi^-$  pair increases exponentially when  moving from the point $a$ to infinity (more precisely, $a$  to  $\infty e^{ i \theta^*} $ where  $\theta^*$ is the phase of corresponding segment)   along the Stokes curve.  The parts between the  Stokes lines is called  Stokes regions or just regions.

When the index of the corresponding Stokes curve is $+$,  $\psi^+$ increases exponentially \footnote{Note that, because of square-root, if   $\psi^+$  increases in first Riemann sheet, it decreases in the second Riemann sheet. This point requires some care at various points. }
 and 
\begin{align}
	\Re \frac{1}{\hbar}\int_a^x \sqrt{Q(x)} dx >0 \,.
\end{align}
When  the index is $-$, then  $\psi^-$ increases exponentially in the  case
\begin{align}
	\Re \frac{1}{\hbar}\int_a^x \sqrt{Q(x)} dx <0 \,.
\end{align}
The Stokes curve indicates where the Stokes phenomena occur,  when $\psi(x)$ is   analytically connected between adjacent 
Stokes regions.

\subsection{Connection formula and monodromy matrix}

For a generic potential and at a typical  value of the energy, the turning points are non-degenerate. To each such  turning point,   one attaches an Airy-type Stokes graph.   Therefore, a Stokes graph for a general potential is a composite of  elementary building blocks of  Airy-type Stokes graphs.  

We now  give a connection formula for the Airy-type Stokes graph.
When one considers the Borel-resummed wave functions for a given potential and its analytic continuation in terms of a complex $x$, 
one has to take into account  the effect of the Borel singularity on the Borel plane, i.e., the Stoke phenomenon.

Roughly speaking, in order to compute the effect,   we decompose a global Stokes graph into the Airy-type Stokes graphs locally and then consider the effect of crossing the Stokes line by using a connection formulas.  
 As we emphasize, the Airy-type Stokes graph is a building block of any given graph, hence, it is important to understand it fully in simple examples.  Apart  from the  connection formula,  one also pay attention  to  the change of normalization point  of the wave function corresponding to the change of turning points. We describe both below. 

\begin{figure}[t]
	\centering 
	\includegraphics[width=9cm]{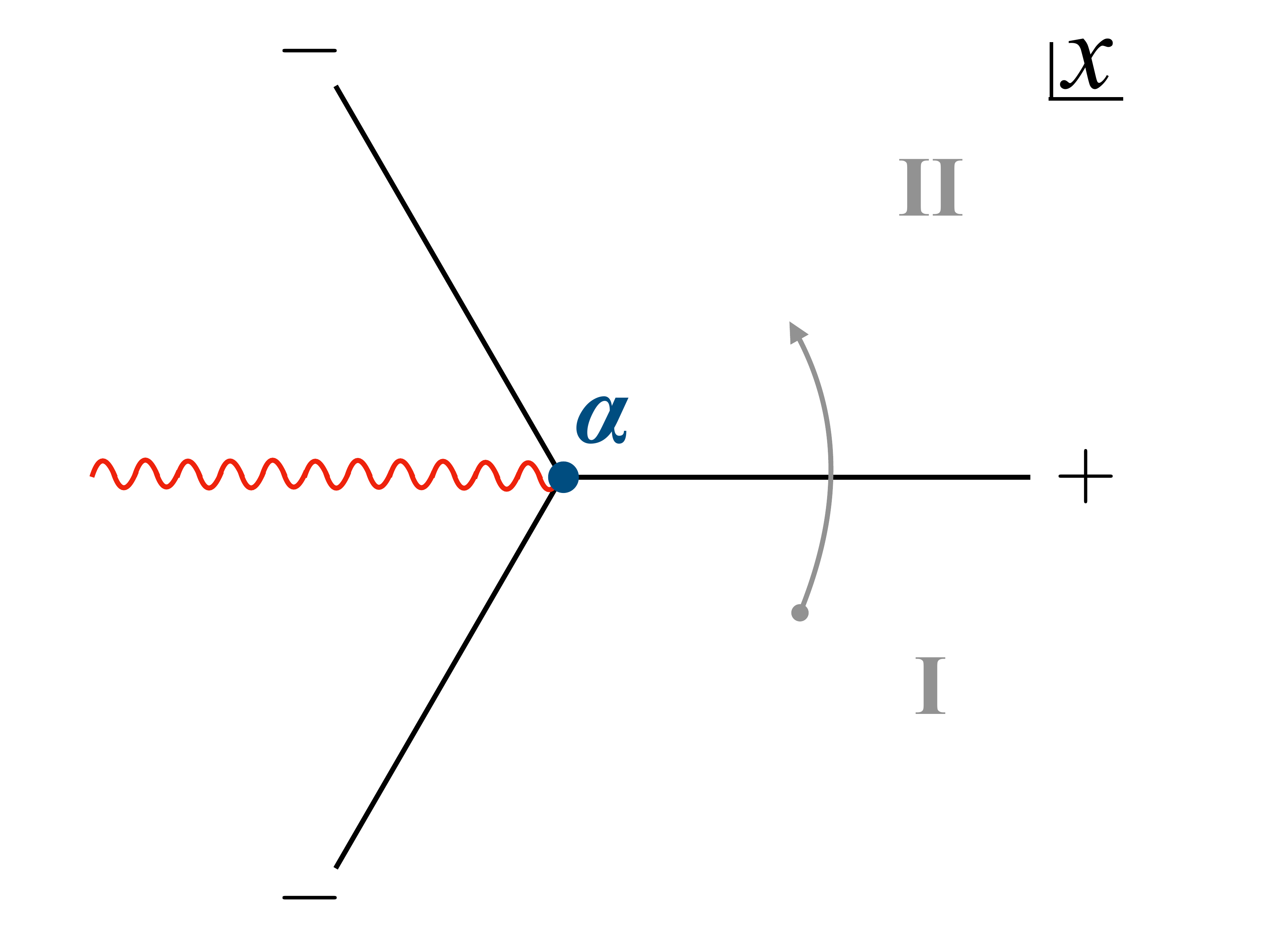}
	\caption{
          The Airy-type Stokes graph emerging from a turning point $a$.
          The sign $+(-)$ labeling each lines means increasing (decreasing) ${\rm Re}\, \frac{1}{\hbar} \sqrt{Q(x)} dx$ as going out from the turning point along the line.  The wavy line  denotes a branch cut. 
          By crossing the curve labeled by $+$  in anti-clockwise manner,   the wave funtions in the ${\rm I}$ and ${\rm II}$ domains are related to each other as $\psi_{a,{\rm I}}=M_+\psi_{a,{\rm II}}$.
        }
        \label{fig:airy}
\end{figure}

We suppose the wave function is normalized at a simple turning point $a$ and consider analytic continuation from the region I to II as shown in Fig.~\ref{fig:airy}. 
When $x$ crosses a Stokes line,  the relation between wave function  can be expressed by
\begin{align}
	\mqty(\psi^{+}_{a,{\rm I}} \\\psi^{-}_{a,{\rm I}})=M\mqty(\psi^{+}_{a,{\rm II}}\\\psi^{-}_{a,{\rm II}}),
\end{align}
The monodromy matrix $M$  multiplies  the wave function according to following rules:  
\begin{align}
  M=
  \begin{cases}
    \mqty(1 &   & i \\ 0 && 1) =:M_+ & \mbox{for anti-clockwise  crossing  of a curve  labeled by $+$} \\
    \mqty(1 & -i \\ 0 & 1) =:M_+^{-1} & \mbox{for clockwise  crossing of a curve  labeled by $+$} \\
    \mqty(1 &   & 0 \\ i && 1) =: M_- &  \mbox{for anti-clockwise  crossing  of a curve labeled by $-$} \\
    \mqty(1 & 0 \\ -i & 1)=: M^{-1}_- &  \mbox{for clockwise crossing of a curve labeled by $-$} 
  \end{cases} 
\end{align}
Furthermore, if it crosses the branch cut emerging from the simple turning point $a$, 
it moves between the first and second Riemann sheets as $S_{\rm odd}(x) \rightarrow -S_{\rm odd}(x)$.
Thus, we have
\begin{align}
M=
  \begin{cases}
    \mqty(0 &   & i \\ i && 0) =:M_b & \mbox{for anti-clockwise  crossing of a cut} \\
    \mqty(0 & -i \\ -i & 0) =:M_b^{-1} & \mbox{for clockwise  crossing of a cut} \\
\end{cases}
\end{align}
A complete cycle around a  turning point gives  identity  map: 
\begin{align}
  M_-M_bM_-M_+=M_+M_bM_+M_-  =\mqty(1 &   & 0 \\0 && 1)\,.
\end{align}
In order to consider the analytic continuation globally beyond the Airy-type Stokes curve, one has to incorporate the change of normalization point and then employ the connection formula due to curves or cuts emerging from other simple turning points.

Two wave functions normalized at  different turning points $a_1, a_2$ are  are related by the equation
\begin{align}
	\psi^\pm_{a_1}(x)=e^{\pm\int_{a_1}^{a_2}S_{{\rm odd}}}\psi^\pm_{a_2}(x) \,.
\end{align}
The  quantity, $\int_{a_1}^{a_2}S_{{\rm odd}}$ is called the Voros multiplier. One may be tempted to think that the Voros multiplier is 
an asymptotic function,   because $S_{{\rm odd}}$ is defined by the recursive relation and asymptotic itself. However, the Voros multiplier appearing here  is    Borel resummed   because the $\psi^\pm_{a_1}(x)$ are already   Borel resummed wave functions. 
We then write down the normalization matrix as
\begin{align}
	\mqty(\psi^+_{a_1}(x) \\ \psi^-_{a_1}(x))&=N_{a_1 a_2}\mqty(\psi^+_{a_2}(x) \\ \psi^-_{a_2}(x)), \qquad 
	N_{a_1 a_2}  =\mqty(e^{+\int_{a_1}^{a_2}S_{{\rm odd}}} & 0 \\ 0 & e^{-\int_{a_1}^{a_2}S_{{\rm odd}}})\,.
\end{align}
 The orientation of $N_{a_1 a_2}$ is flipped before/after crossing a branch cut.
\be
M_{b} N_{a_1 a_2} = N^{-1}_{a_1 a_2} M_{b}\,.
\ee

\subsection{Warm-up: Harmonic oscillator with Airy-type Stokes graph}\label{sec:HO}

%
%
%
%
%

\begin{figure}[t]
	\centering 
  \includegraphics[width=6cm]{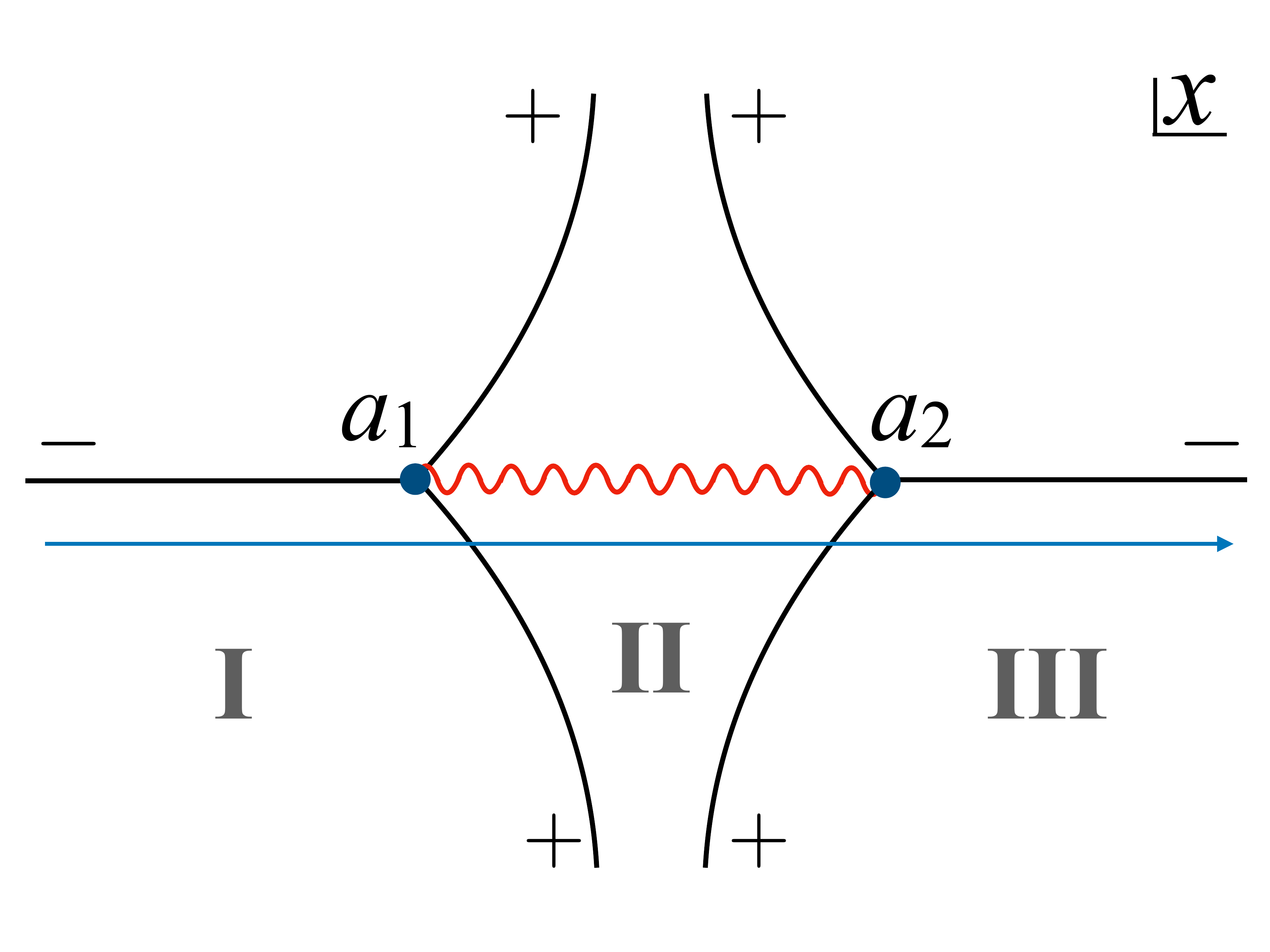}
  \quad\quad
  \includegraphics[width=6cm]{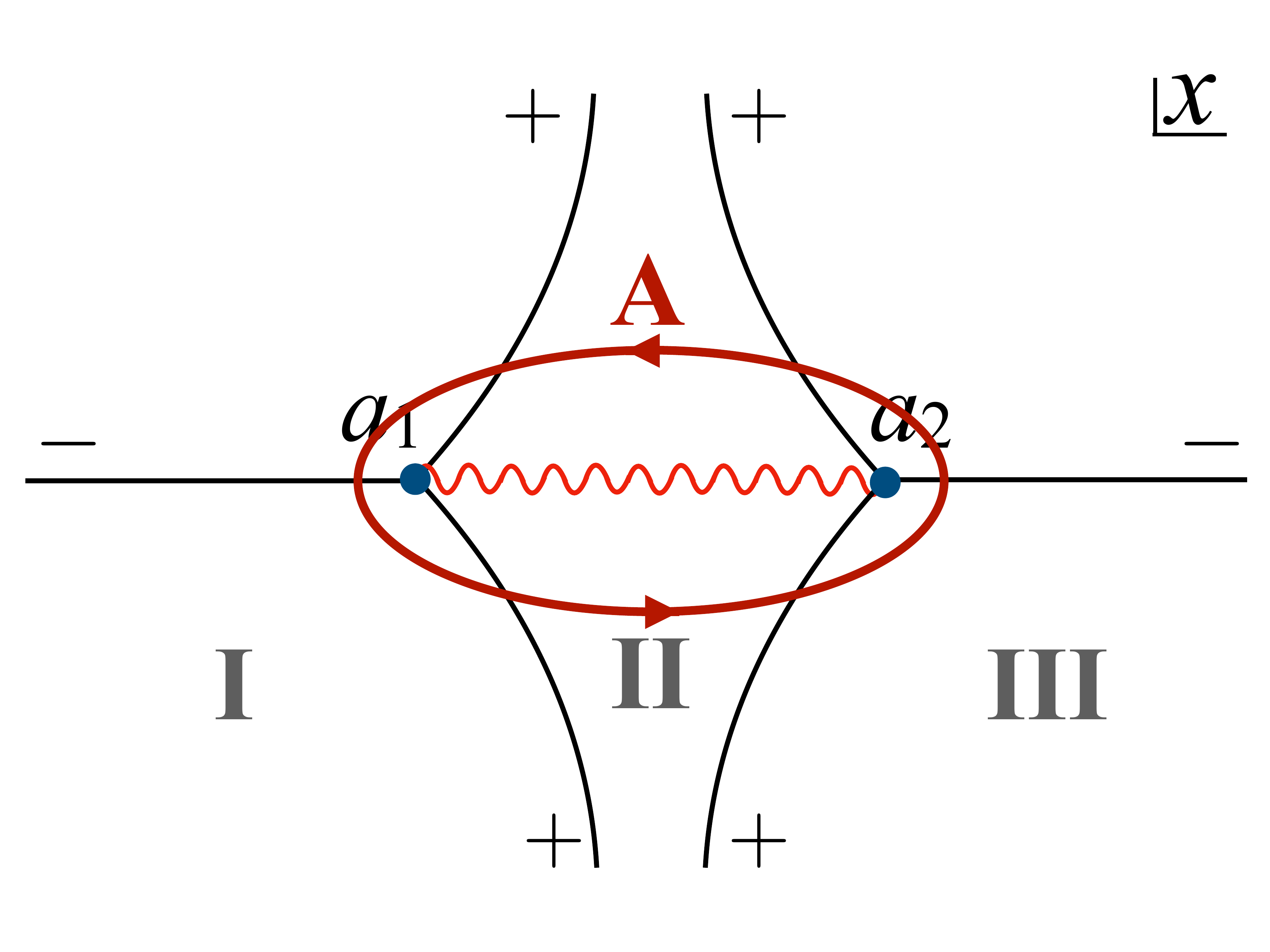}
  \caption{
    The Stokes graph for harmonic oscillator.
    In order to obtain the quantization condition, we take the orbit from the left to the right below the real axis by  taking into account the Stokes phenomena (Left panel).
    The cycle $A$ is defined as an oriented cycle enclosing two turning points $a_1$ and $a_2$ (Right panel).
    }
  \label{Harmonic_cycle}
\end{figure}

For a general potential  with multiple degenerate harmonic minima, we associate a Stokes graph which is a combination  of the  Stokes graph of harmonic oscillator.     The Stokes graph for harmonic oscillator is a combination of two Airy-type graphs as shown in 
Fig.\ref{Harmonic_cycle}.  Therefore, one can quickly learn how  the formalism works in practice in this simple, but essential example. 
Therefore, we first review this example, and then  move to more interesting examples of double-well, triple-well  and $N$-ple well examples. 

The harmonic  potential is given by  $V(x)=\frac{1}{2}\omega^2x^2$. Its Stokes curve is depicted in Fig.~\ref{Harmonic_cycle} assuming $E>0$.   
There are two turning points,  $a_1=-\frac{\sqrt{2E}}{\omega}$ are $a_2=\frac{\sqrt{2E}}{\omega}$,  which satisfy $Q(x)=2(V(x)-E)=0$.  
The blue arrow is a trajectory of the analytic continuation. If we start with a  decaying solution at the beginning of blue line, we will demand a decaying solution at the end of journey, for the full WKB solution to be normalizable.  However, the Stokes phenomena will induce terms that will  be exponentially growing at the end. In order to have a  physical answer, we will demand that the pre-factor of the  exponentially growing part to vanish. That will give us the quantization condition that will determine the spectrum of the theory.\footnote{There are several other ways to obtain the quantization condition using exact-WKB method. For example, the Wronskian constraint for each Stokes region is used in \cite{Ito:2018eon}.}

First,  let us take wave function normalized at $a_1$ and consider analytic continuation from the region I to I\hspace{-.1em}I. Then the wave function changes as
\begin{align}
	\mqty(\psi^+_{a_1,\textrm{I}}(x) \\ \psi^-_{a_1,\textrm{I}}(x))=M_+\mqty(\psi^+_{a_1,\textrm{I\hspace{-.1em}I}}(x) \\ \psi^-_{a_1,\textrm{I\hspace{-.1em}I}}(x))\,.
\end{align}
Second,  consider analytic continuation from the region I\hspace{-.1em}I to I\hspace{-.1em}I\hspace{-.1em}I. In this case, the Stokes curve to be crossed is starting from the other turning point, $a_2$. Therefore we have to change the normalization as
\begin{align}
	\mqty(\psi^+_{a_1,\textrm{I\hspace{-.1em}I}}(x) \\ \psi^-_{a_1,\textrm{I\hspace{-.1em}I}}(x))=N_{a_1a_2}\mqty(\psi^+_{a_2,\textrm{I\hspace{-.1em}I}}(x) \\ \psi^-_{a_2,\textrm{I\hspace{-.1em}I}}(x))\,.
\end{align}
Then we can multiply the monodromy matrix as
\begin{align}
	\mqty(\psi^+_{a_2,\textrm{I\hspace{-.1em}I}}(x) \\ \psi^-_{a_2,\textrm{I\hspace{-.1em}I}}(x))=M_+\mqty(\psi^+_{a_2,\textrm{I\hspace{-.1em}I\hspace{-.1em}I}}(x) \\ \psi^-_{a_2,\textrm{I\hspace{-.1em}I\hspace{-.1em}I}}(x))\,.
\end{align}
As a result,\footnote{One can omit the last $N_{a_2 a_1}$ because it just changes overall factors but this  makes $D$ simpler.}, we obtain the connection formula
\begin{align}
	\mqty(\psi^+_{a_1,\textrm{I}}(x) &   \\ \psi^-_{a_1,\textrm{I}}(x))&=M_+N_{a_1a_2}M_+N_{a_2a_1}\mqty(\psi^+_{a_1,\textrm{I\hspace{-.1em}I\hspace{-.1em}I}}(x) \\ \psi^-_{a_1,\textrm{I\hspace{-.1em}I\hspace{-.1em}I}}(x))\\
	& =\mqty(\psi^+_{a_1,\textrm{I\hspace{-.1em}I\hspace{-.1em}I}}(x)+i(1+A)\psi^-_{a_1,\textrm{I\hspace{-.1em}I\hspace{-.1em}I}}(x) \\\psi^-_{a_1,\textrm{I\hspace{-.1em}I\hspace{-.1em}I}}(x))\,,
\end{align}
where the cycle $A=e^{\oint_A S_{{\rm odd}}}=e^{2\int_{a_1}^{a_2}S_{{\rm odd}}}$ is depicted 
in Fig.~\ref{Harmonic_cycle}.

As $x\rightarrow -\infty$ in the region $\textrm{I}$, $\psi^+$ is normalizable, it decays as $x\rightarrow -\infty$. (This is true on first Riemann sheet which we stick through this argument.)
Therefore we take $\psi^+$ in the region $\textrm{I}$ and we find that it changes to $\psi^+_{a_1,\textrm{I\hspace{-.1em}I\hspace{-.1em}I}}(x)+i(1+A)\psi^-_{a_1,\textrm{I\hspace{-.1em}I\hspace{-.1em}I}}(x)$ in the region $\textrm{I\hspace{-.1em}I\hspace{-.1em}I}$. 
In region  $\textrm{I\hspace{-.1em}I\hspace{-.1em}I}$,  $\psi^+_{a_1,\textrm{I\hspace{-.1em}I\hspace{-.1em}I}}(x)$ is decaying as 
 $x\rightarrow -\infty$ while  $\psi^-_{a_1,\textrm{I\hspace{-.1em}I\hspace{-.1em}I}}(x)$  is blowing up. Therefore, 
 in order to satisfy the normalization condition, the coefficient of $\psi^-$ {\bf must} be zero. This is the quantization condition:
\begin{align}
  D=1+A=1+e^{\oint_A S_{{\rm odd}}}=0 \,.
\end{align}
where $e^{\int_{a_1}^{a_2}S_{{\rm odd}}}$ is the Voros multiplier connecting two turning points. 
 This is equivalent to
\begin{align}
	\oint_A S_{{\rm odd}}=-2\pi i\qty(n+\frac{1}{2})\;\;\;\text{with }n\in\mathbb{Z} \,.
\end{align}
For the  harmonic oscillator,  
\begin{align}
\oint_A S_{{\rm odd}}  = \oint_A S_{{-1}}  =     \frac{1}{\hbar}\oint_A \sqrt{2(V(x)-E)}dx   =-2\pi i\frac{E}{\hbar\omega} \,,
\end{align}
where $\omega=\frac{2\pi}{T}$ and  $T$ is the classical period,  $T=\qty|\oint_A ({2(V(x)-E)} )^{-1/2}dx|$.  
Therefore,  the quantization condition $D(E)=0$ obtained from the exact-WKB analysis gives
\begin{align}
E=\hbar\omega\qty(n+\frac{1}{2})\,.
\end{align}
The Stokes curve in Fig.\ref{Harmonic_cycle} corresponds to $E>0$ and the turning points are real. This puts a restriction that $n=0,1,2...$, which is just  
 the spectrum of simple  harmonic oscillator.  
%

\subsection{Resolvent and Spectral form} 
\label{sec:resolvent}
We derive the     partition function starting with  the quantization condition $D(E)$  and resolvent $G(E)$.  The reason we are presenting this  is  because we will follow verbatim the same procedure in the  theories  with instantons and we will reach to fairly non-trivial results. It is therefore useful to recall this  tool in a simple example. 

\begin{figure}[H]
	\centering 
	\includegraphics[width=8cm]{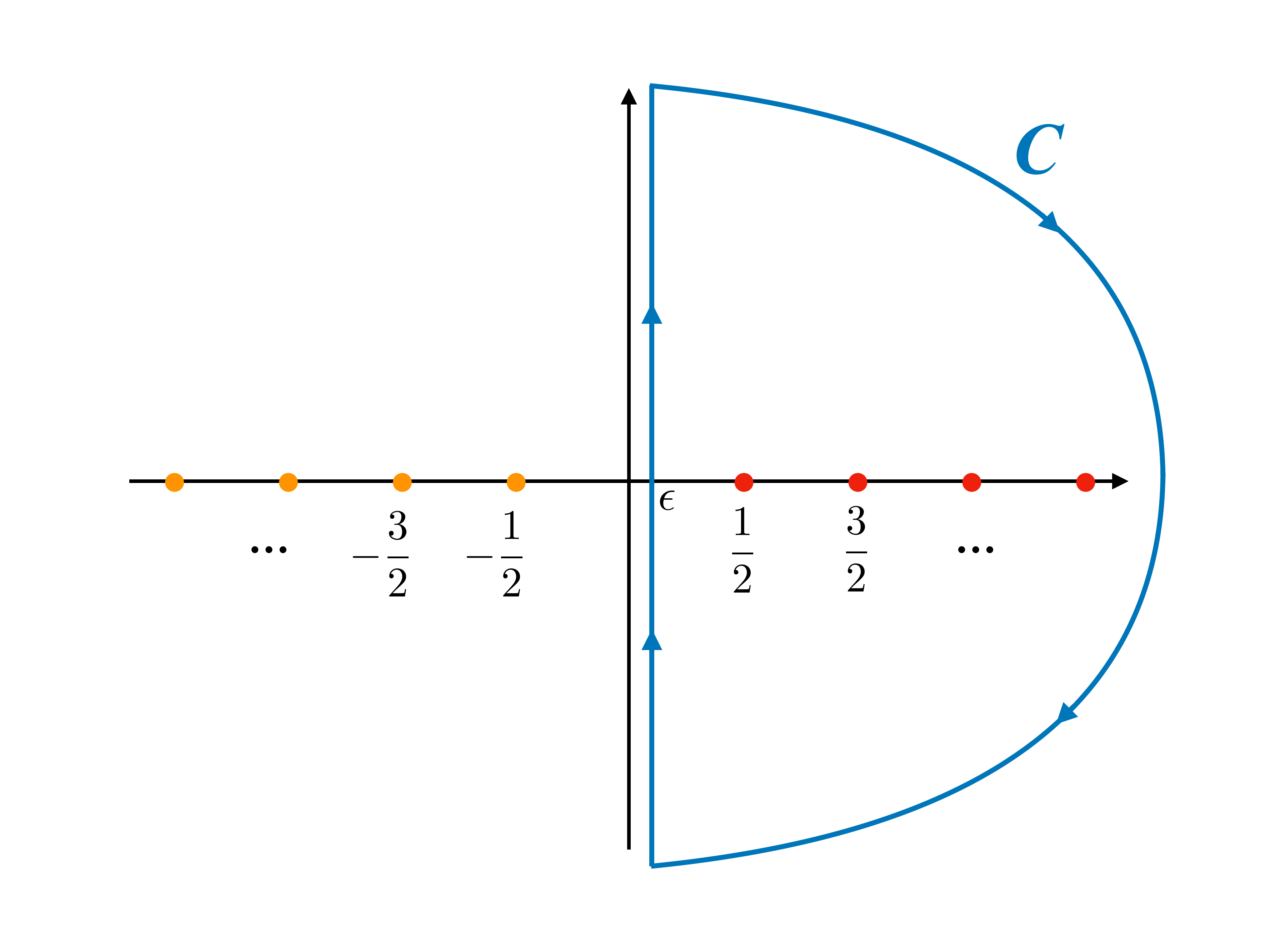}
	\caption{$C$ is the   integration contour in the determination of  the partition function as the inverse Laplace transform of resolvent. 
}
          \label{fig:contour}
\end{figure}

The quantization condition $D$ is written as \footnote{ The  zeta function regularization for Fredholm determinant $D(E)=\det(H-E)$ gives: 
\begin{align}
	e^{-\zeta'(0,\frac{1}{2}-\frac{E}{\hbar\omega})}=\frac{\sqrt{2\pi}}{\Gamma\qty(\frac{1}{2}-\frac{E}{\hbar\omega})} 
\end{align}
It removes the irrelevant Gamma function, which does not contribute to the partition function defined through contour $C$.
} 
\begin{align}
	D & =1+e^{-2\pi i\frac{E}{\hbar\omega}} =e^{-\pi i\frac{E}{\hbar\omega}}2\sin\qty(\pi\qty(\frac{E}{\hbar\omega}+\frac{1}{2}))                                           \nonumber\\
	  & =e^{-\pi i\frac{E}{\hbar\omega}}\frac{2\pi}{\Gamma(\frac{1}{2}+\frac{E}{\hbar\omega})\Gamma(\frac{1}{2}-\frac{E}{\hbar\omega})} \,,
\end{align}
where we have used reflection formula: $\Gamma(x)\Gamma(1-x)=\frac{\pi}{\sin(\pi x)}$.
We then obtain the resolvent $G(E)$ from the quantization condition as:
\begin{align}
	G(E) & =-\pdv{E}\log D  \label{eq:G_logD} \nonumber\\
	     & =\frac{\pi i}{\hbar\omega} +\pdv{E}\log\Gamma\qty(\frac{1}{2}+\frac{E}{\hbar\omega})+\pdv{E}\log\Gamma\qty(\frac{1}{2}-\frac{E}{\hbar\omega})\,.
\end{align}
The partition function is the inverse Laplace transform of resolvent:
\begin{align}
	Z=\frac{1}{2\pi i}\int_{\epsilon-i\infty}^{\epsilon+i\infty}G(E)e^{-\beta E}dE \,.
\end{align}
To calculate this quantity, we consider the contour $C$ depicted in Fig.~\ref{fig:contour},
where $C$ is determined by the condition $E>0$. 
It leads $-\frac{1}{2}\hbar\omega<\epsilon<\frac{1}{2}\hbar\omega$ and $C$ is closing  in the positive real region.

Inside  the contour $C$, $\frac{\pi i}{\hbar\omega}$ and $\pdv{E}\log\Gamma\qty(\frac{1}{2}+\frac{E}{\hbar\omega})$ are holomorphic, hence do not contribute to integration.  Furthermore, $\pdv{E}\log\Gamma\qty(\frac{1}{2}-\frac{E}{\hbar\omega})$ has infinitely many poles in $C$ and all the residues\footnote{$\pdv{x}\log\Gamma(x)\sim\frac{1}{x}$ in $x\rightarrow 0$} are 1.
Therefore we find
\begin{align}
	Z(\beta)=\frac{1}{2\pi i}\int_C\qty[\pdv{E}\log\Gamma\qty(\frac{1}{2}-\frac{E}{\hbar\omega})]e^{-\beta E}dE=\sum_{n=0}^\infty e^{-\beta\hbar\omega\qty(n+\frac{1}{2})} \,.
\end{align}
the partition function of harmonic oscillator.  We will use the same strategy in more general cases involving instantons  to describe the 
 partition functions of the systems  from the exact quantization condition.


\vspace{-0.5cm}
\section{Symmetric double-well potential} 
\label{sec:DW}

We consider the exact-WKB analysis for  the symmetric double-well potential. It is known that (1) the leading non-perturbative contribution to its ground state energy comes from the instanton configuration, and (2) the Borel ambiguity of the  perturbation  theory for the ground state  is cancelled by that of the bion (correlated instanton--anti-instanton configuration) contribution.  This  pattern continue to higher states under the barrier. 
The exact form of the bion  contribution can be obtained  from the   quasi-zero mode integration (quasi-moduli integral).
 We first review  the resurgent structure of the partition function in  this system. 
 Then, we find an explicit mapping between this construction and Gutzwiller's quantization.
In doing so,  we  figure out the  relation between the phase ambiguity of quasi-moduli integral,  the topological properties of the Stokes curve in terms of Gutzwiller's quantization.

\begin{figure}[H]
	\centering 
	\includegraphics[width=6cm]{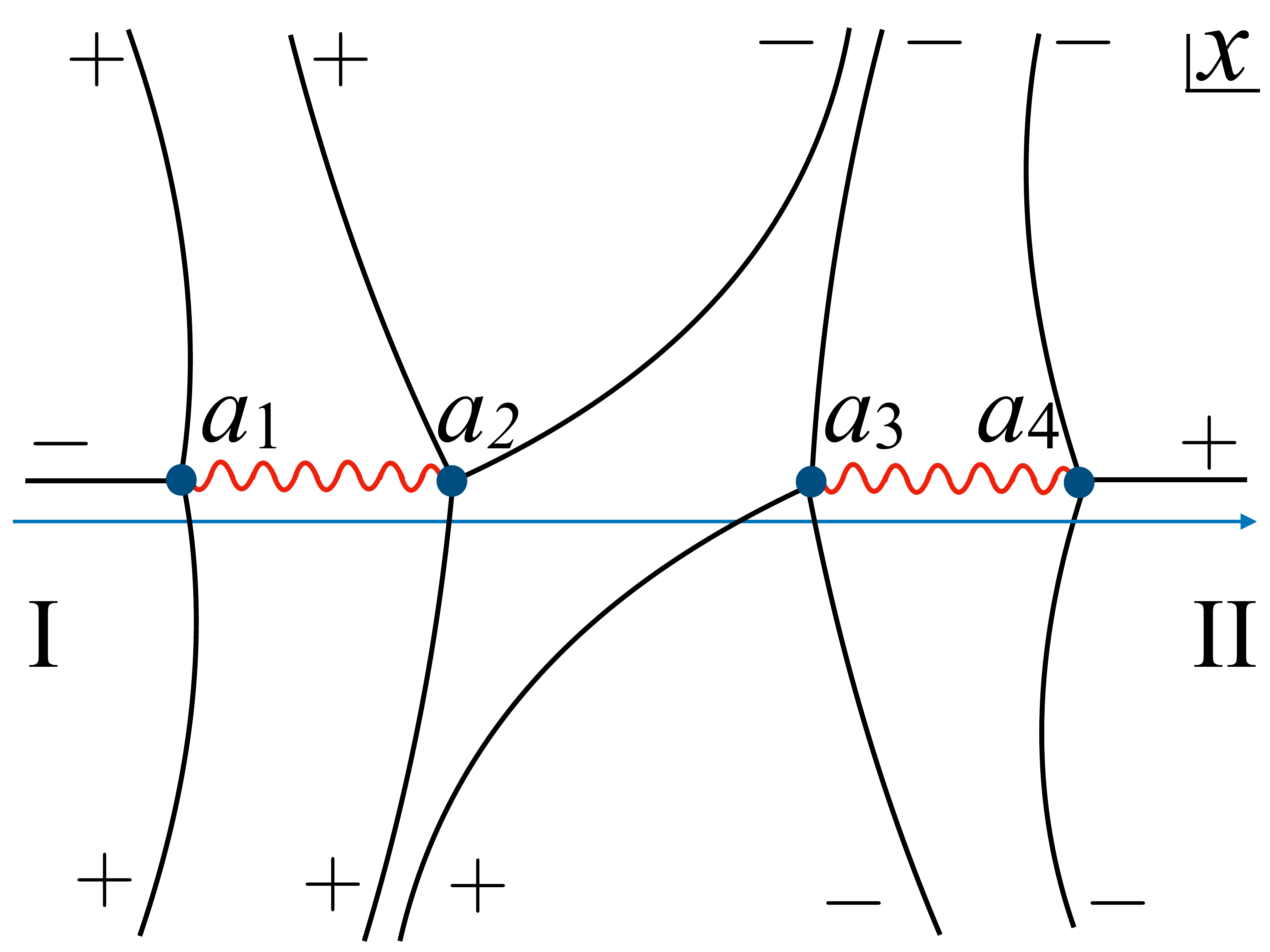}\quad\quad
	\includegraphics[width=6cm]{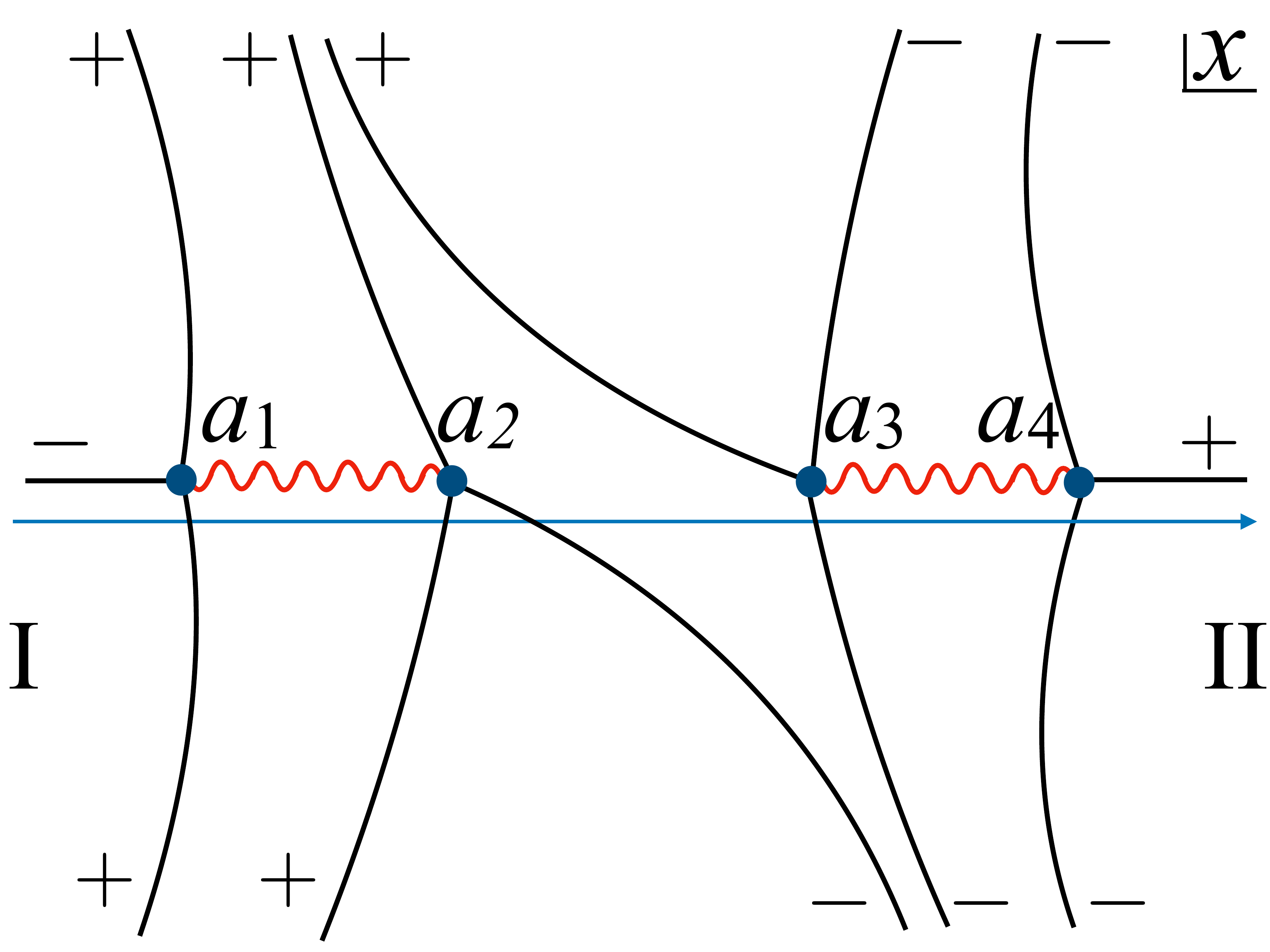}
  \caption{
  The Stokes graph for the double-well potential with $\Im \hbar > 0$ (Left panel) and  $\Im\hbar < 0$ (Right panel).
  We took the two branch cuts such that their end-points are turning points $(a_1,a_2)$ and $(a_3,a_4)$.
  We take the orbit for obtaining the quantization condition from the left to the right below the real contour.
  }
\label{fig:DW12}
\end{figure}

For the  symmetric double-well  potential,  $Q(x) = 2(V(x) -E ) =(x-a_1)(x-a_2)(x-a_3)(x-a_4)$ where $a_i$ are turning points.  Then the Stokes curve of this systems\footnote{This Stokes curve corresponds   to the  low energy region,  to energies below the barrier height. Above the barrier height, two of the real turning points, $a_2$ and $a_3$ turns into complex conjugate turning points.   
Even in high energy region, we can show the topological structure of Stokes curve is corresponding to phase ambiguity. } is schematically depicted as shown in Fig.~\ref{fig:DW12}.

\begin{figure}[H]
	\centering 
	\includegraphics[width=6cm]{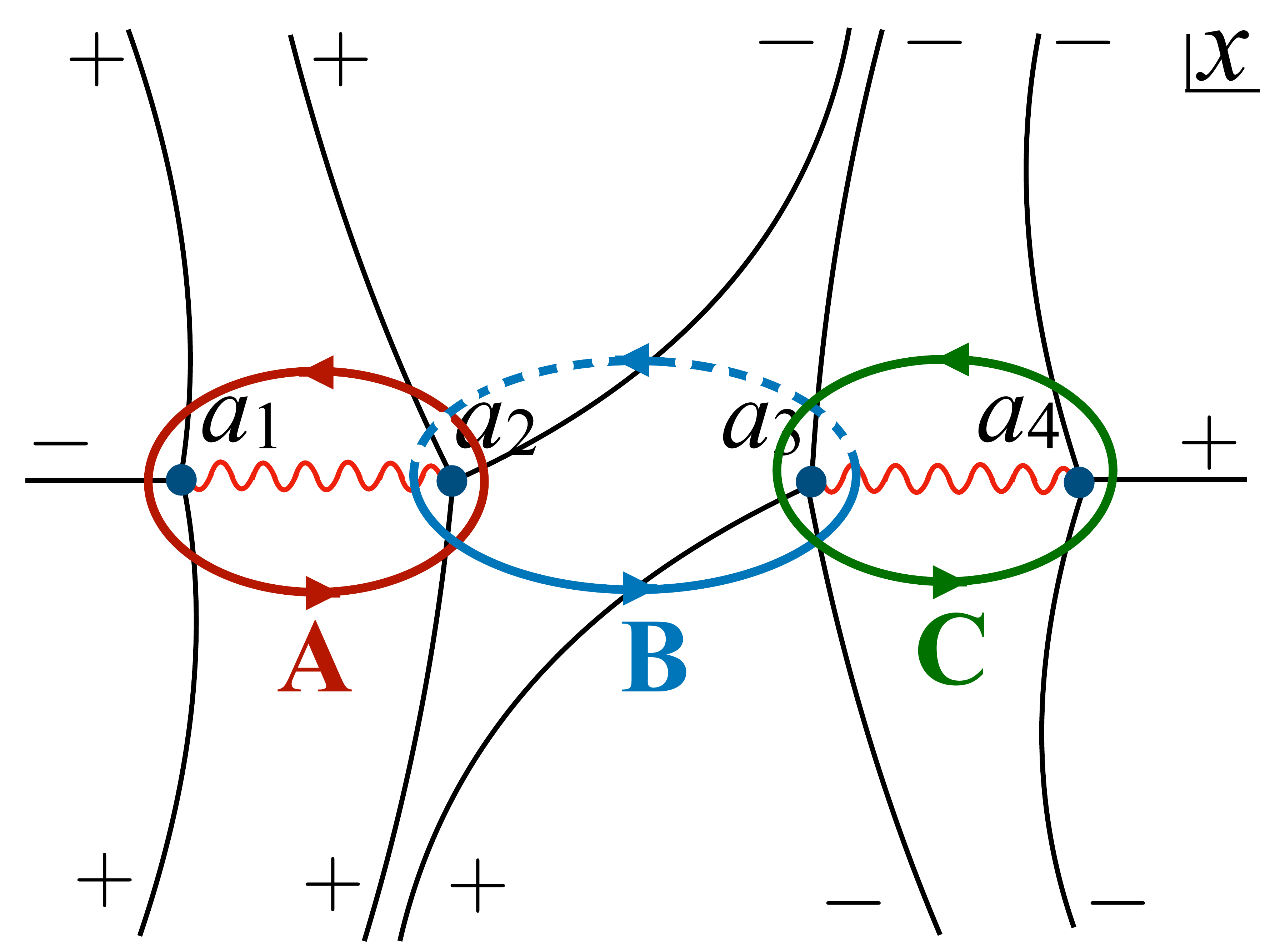}\quad\quad
	\includegraphics[width=6cm]{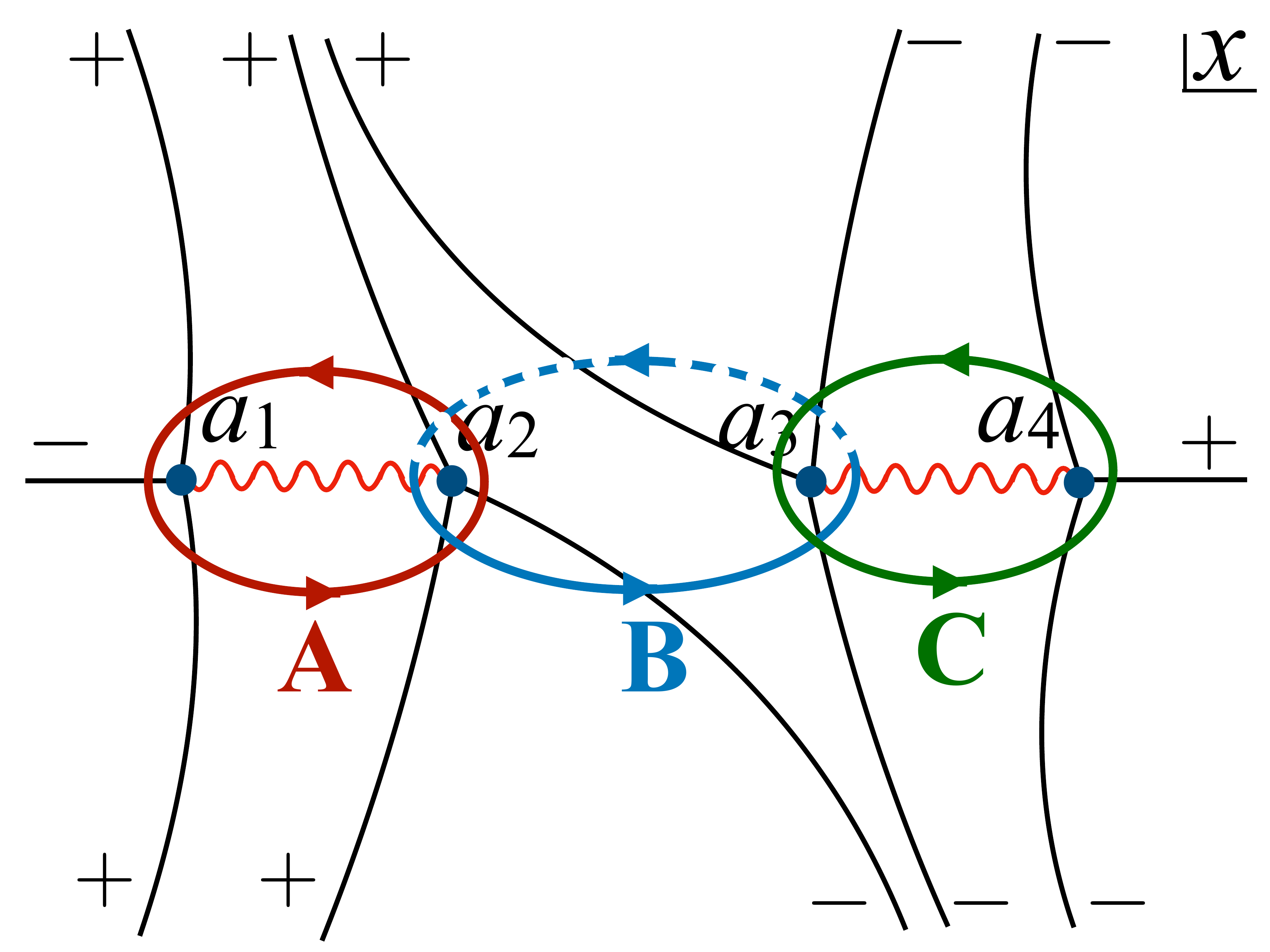}
  \caption{
    Perturbative cycles ($A,C$) and a non-perturbative cycle ($B$) for the symmetric double-well potential.
    The perturbative cycles, $A$ and $C$, are defined as oriented cycles enclosing $(a_1,a_2)$ and $(a_3,a_4)$, respectively, and the non-perturbative cycle, $B$, is an oriented cycle enclosing $(a_2,a_3)$ and intersecting with the two branch cuts.
  }
\label{DoubleWell_cycle}
\end{figure}

As we did for the harmonic oscillator in Sec.~\ref{sec:HO}, we obtain the quantization condition from the normalized condition by performing the analytical continuation of the wave function from $-\infty$ to $\infty$ using the connection formulas.
\begin{align}
\begin{cases}
\mqty(\psi^+_{a_1,\textrm{I}}(x) \\ \psi^-_{a_1,\textrm{I}}(x))=M_+N_{a_1a_2}M_+N_{a_2a_3}M_+M_-N_{a_3a_4}M_-N_{a_4a_3}N_{a_3a_2}N_{a_2a_1}\mqty(\psi^+_{a_1,\textrm{II}}(x) \\ \psi^-_{a_1,\textrm{II}}(x))\quad\quad \mbox{for \ $\Im \hbar>0$}  \\
\\
\mqty(\psi^+_{a_1,\textrm{I}}(x) \\ \psi^-_{a_1,\textrm{I}}(x))=M_+N_{a_1a_2}M_+M_-N_{a_2a_3}M_-N_{a_3a_4}M_-N_{a_4a_3}N_{a_3a_2}N_{a_2a_1}\mqty(\psi^+_{a_1,\textrm{II}}(x) \\ \psi^-_{a_1,\textrm{II}}(x)) \quad\quad \mbox{for \ $\Im \hbar<0$}
\end{cases}
\end{align}

The quantization condition for this case is given by
\begin{align}
&D \propto
\begin{cases}
(1+A^{+})(1+C^{+})+A^{+} B^{+}=0  \quad\quad \mbox{for \ $\Im \hbar>0$}  \\
(1+A^{-})(1+C^{-})+C^{-} B^{-}=0 \quad\quad \mbox{for \ $\Im \hbar<0$}
\end{cases},\label{eq:D_hbp}
\end{align}
where the cycles are defined as\footnote{$C$ is the same as $A$ on the other Riemann sheet (See the index of Stokes curve) since the potential is symmetric.} 
\begin{align}
A=e^{\oint_A S_{{\rm odd}}},   \qquad B=e^{\oint_B S_{{\rm odd}}},  \qquad  C=e^{\oint_C S_{{\rm odd}}}=1/A 
\end{align}
and ${\frak C}^{\pm}:= {\cal S}_{\pm}[{\frak C}]$ for ${\frak C}\in \{A,B,C\}$ as shown in Fig.~\ref{DoubleWell_cycle}. The $A$ and $C$ cycles are perturbative and $B$ cycle is non-perturbative, $B\propto e^{-\frac{S}{\hbar}}$, where $S$ corresponds to the single bion contribution.

We here defined the notation, where ${\frak C}$ is used as series forms and ${\frak C}^{\pm}$ is used as Borel-summed forms.
However, from now on, we would use a simplified notation where we use ${\frak C}$ instead of ${\frak C}^{\pm}$ for simplicity unless it causes a confusion.\footnote{${\frak C}^{\pm}$ can be identified when we look only at its exponentially dominant sector asymptotically.}

To evaluate the non-perturbative contribution to the ground state energy and the phase ambiguity term, 
let us consider the asymptotic form of $A$, which does not include non-perturbative contribution,  before being Borel-resummed. It is: 
\begin{align}
  A \rightarrow e^{-2\pi i \frac{E}{\hbar \omega_A(E,\hbar)}}\,.
  \label{Asymptotic_omega}
\end{align}
This $\omega_A(E,\hbar)$ is an  asymptotic expansion in $\hbar$.  In the low energy limit, it  can be regarded as a harmonic frequency of the classical (harmonic)  vacuum: 
\begin{align}
  \omega_A(E,\hbar)^2&=\sum_{n=0}^\infty c_n(E)\hbar^n\\
  \lim_{E\rightarrow 0}c_0(E)&=V''(x_{\rm vac})\,,
\end{align}
where $x_{\rm vac}$ is a minimum  of the potential. 
We emphasize that writing down this expression corresponds to taking the Borel-resummed $A$ back to its  asymptotic expansion form. This procedure helps us to see that the quantization condition $D$ has the phase ambiguity. However, of course, this ambiguity disappears when we consider the Borel-resummed form. 
We will show it in the next subsection.

We now set $E=\hbar \omega_A \qty(\frac{1}{2}+\delta)$, where $\delta$ roughly stands for the energy deviation from that of the harmonic oscillator. 
The quantization condition $D=0$ then becomes 
\begin{align}
  &4\sin^2(\pi\delta)=e^{-2\pi i\delta}B\quad\;\;\;\;\;\Im \hbar>0\,,\nl
  &4\sin^2(\pi\delta)=e^{2\pi i\delta}B\quad \;\;\;\;\;\Im \hbar<0\,.
\end{align}
Or equivalently, 
\begin{align}
  \frac{1}{\Gamma(-\delta)} & =\pm\frac{\sqrt{B}}{2\pi}e^{-\pi i\delta}\Gamma(1+\delta)\quad\;\;\;\;\; \Im \hbar>0\,,\nl
	\frac{1}{\Gamma(-\delta)} & =\pm\frac{\sqrt{B}}{2\pi}e^{\pi i\delta}\Gamma(1+\delta)\quad\;\;\;\;\; \Im \hbar<0\,.
	\label{eigenvalue_D}
\end{align}
Here $\pm$ in the latter form stands for parity. We emphasize that this result is obtained without any approximation. 

In \cite{ZinnJustin:1981dx}\cite{ZinnJustin:2004ib},\cite{Sueishi}, The quantization condition was calculated using path integral(QMI) method. The result is
\begin{align}
  \frac{1}{\Gamma(-x)}&=\pm \frac{e^{-S_{inst}}}{2\pi}e^{-\pi ix}\qty(\frac{\hbar}{2})^{-x-\frac{1}{2}}\sqrt{2\pi}\quad\;\;\;\;\; \Im \hbar<0\,.\nl
  \frac{1}{\Gamma(-x)}&=\pm \frac{e^{-S_{inst}}}{2\pi}e^{\pi ix}\qty(\frac{\hbar}{2})^{-x-\frac{1}{2}}\sqrt{2\pi}\quad\;\;\;\;\; \Im \hbar<0\,.,
  \label{eigenvalue_P}
\end{align}
where $x=E-\frac{1}{2}$.
Considering that $\qty(\frac{\hbar}{2})^{-\delta-\frac{1}{2}}\sqrt{2\pi}$ in (\ref{eigenvalue_P}) is the contribution from quantum fluctuations, this part is included in $B$ and $\omega_A$ in (\ref{eigenvalue_D}). The extra Gamma function $\Gamma(1+\delta)$ is coming from the negative energy part when we consider the argument in (\ref{eq:G_logD}), so it can be ignored under the condition $E>0$. Therefore, this result is regarded as the complete quantization condition with full quantum fluctuations.




\subsection{Gutzwiller's quantization}
\label{sec:Gutzwiller_quantization}

Gutzwiller's quantization is based on prime-periodic orbit (p.p.o.) as a fundamental unit, but the way how to add up this p.p.o. has not been clearly known except for simple systems. 
We will see that one can exactly obtain the Gutzwiller's form from the quantization conditions in the exact-WKB analysis and it reveals a new physical meaning of the quasi-moduli integral in the path integral method. 

First,  let us rewrite the quantization condition Eq.~\eqref{eq:D_hbp}, using $C= 1/A$, in terms of only $A$ and $B$ cycles 
\begin{align}
  D(E)=  (1+A)(1+A^{-1} ) \left ( 1 +  \frac{  B}{ D_A^2}  \right)
  \label{eq:quancon}
\end{align}
where $D_A$=$1+A^{-1}\,\, (\Im\hbar>0)$ or $1+A \,\,(\Im\hbar <0)$.   This 
rewriting allows us to write  the trace of resolvent $G(E)=-\pdv{E}\log D(E)$,   
derived from the quantization condition  Eq.~(\ref{eq:D_hbp}),  in a useful form: 
\begin{align}
	G(E) & = G_{\rm p}(E)+G_{\rm np}(E) \nonumber\\
	&=\qty[-\pdv{E}\log (1+A)-\pdv{E}\log (1+A^{-1})]+\qty[-\pdv{E}\log\qty(1+\frac{B}{D_A^2})]\nonumber\\
	&=\qty[-\frac{\pdv{E}A}{1+A}-\frac{\pdv{E}A^{-1}}{1+A^{-1}}]+\qty[-\frac{\pdv{E}(D_A^{-2} B)}{1+(D_A^{-2} B)}],
	\label{resolvent-1}
	\end{align}
The derivative term $\pdv{E}A$ produces the ``period''
\begin{align}
\pdv{E}A & =\pdv{E}e^{\oint_A S_{{\rm odd}}}      =\qty(\pdv{E}\oint_A S_{{\rm odd}})e^{\oint_A S_{{\rm odd}}}                                 \nonumber\\
& =\qty(\oint_A\frac{1}{\hbar}\frac{-1}{\sqrt{2(V-E)}}+O(\hbar))e^{\oint_A S_{{\rm odd}}}  \equiv-\frac{1}{\hbar}iT_{A}A\,.                                                    
\end{align}
and similarly, 
\begin{align}
\pdv{E}B & = -\frac{1}{\hbar}iT_{B}B\,.                                                    
\end{align}
\begin{align}
\pdv{E} \qty(D_A^{-2}B)  & = -i\frac{1}{\hbar} \sum_{n, m=1}^\infty (-1)^{(n+m)} \big( T_B \mp (n+m) T_A  \big) B(A^{\mp})^{n+m},                                                  
\end{align}
where $\pm$ corresponds to $\Im \hbar>0$ and $\Im\hbar<0$ respectively. Since the classical solutions in the lower part of the potential are doubly-periodic, and our definition of $S_{-1} = \sqrt Q $ where $Q=2(V-E)$, $T_A$ is real ($E>V$) and $T_B$ is purely imaginary ($E<V$). Our construction instructs us that {\it complex} periodic paths are part of Gutzwiller formula.
This seems to be the mechanism through which Gutzwiller formula is able to capture the tunneling (instanton) effects. 
The  magnitudes of the quantities  $T_A, T_B$ can be called   {\it quantum periods} and its leading term corresponds exactly to the period of the classical orbit. 
Using these quantities, $G(E)$ can be expressed as
\begin{align}
    G(E)&=G_{\rm p}+G_{\rm np}\\
	G_{\rm p}(E)&=i\frac{1}{\hbar}T_A\sum_{n=1}^\infty(-1)^nA^n+i\frac{1}{\hbar}T_A\sum_{n=1}^\infty(-1)^nA^{-n},\\
	G_{\rm np}(E)&=-\pdv{E}\qty(D_A^{-2}B)   
	\sum_{n=1}^\infty(-1)^n(D_A^{-2}B)^n, \label{eq:DA2B}\\ 
	D_A^{-2}B&=\begin{cases}
		B\qty(\sum_{k=1}^\infty(-1)^k A^{-k})\qty(\sum_{l=1}^\infty(-1)^lA^{-l}) & (\Im\hbar>0)\\
		B\qty(\sum_{k=1}^\infty(-1)^k A^k)\qty(\sum_{l=1}^\infty(-1)^lA^l) & (\Im\hbar<0)
	\end{cases}
\end{align}
This is exactly the form of Gutzwiller's quantization in Eq.~(\ref{Gutzwiller_form}) including the quantum corrections. Note that the quantum period and each cycle contain the quantum corrections (e.g. $T_A=T_{A,cl}+O(\hbar)$, $ A=e^{\frac{i}{\hbar}\oint_A p}+O(\hbar)$). It is important to remind ourselves that the $(-1)$ associated with each cycle can be interpreted as the factor coming from Maslov index (See Sec.~\ref{sec:AppendixMaslov}). 

\begin{figure}
	\centering 
	\includegraphics[width=8cm]{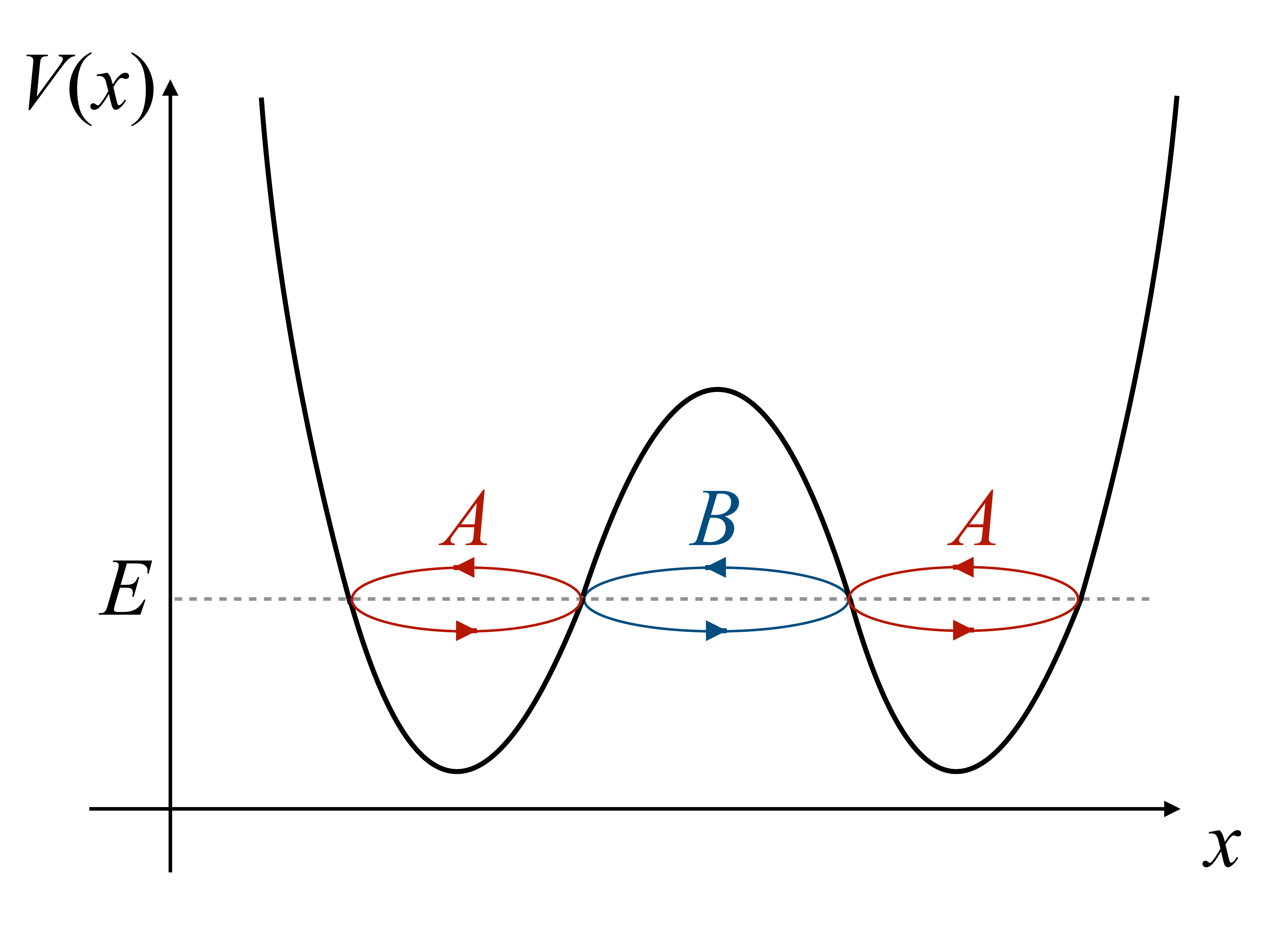}
	\caption{
	Relationship between a periodic orbit and the Maslov index for the symmetric double-well potential.
	The index $(-1)^n$ is determined by counting $D_A^{-2} B$ in Eq.(\ref{eq:DA2B}), as a unit.
	$D_A^{-2}B$ includes two infinite number of $A$-cycles ($D_A^{-1}=\frac{1}{1+A}$) and one $B$-cycle ($B$).
	}
	\label{fig:D_AB}
\end{figure}

 Our result shows what p.p.o. are and how to add them up explicitly, and it is by no means obvious.  Perhaps, we should take 
exact quantization condition and the corresponding resolvent Eq.\eqref{resolvent-1}  as  the precise meaning of the Gutzwiller's sum. 
The perturbative part consists of the infinite number of $A$ cycles and the non-perturbative part is made up of the   infinite number of $A$ cycles and   $B$ cycle.  
 The change of topology of the Stokes curves corresponds to the reversal of the direction of the $A$ cycle of the non-perturbative term.  As we show later, this transition can give the new perspective of the quasi-moduli integration.

\subsection{Partition function} 
\label{sec.part_func}

In this subsection, we calculate the partition function based on the resolvent method in Sec.\ref{sec:resolvent}.
In particular, when evaluating the partition function using path integral, it is important to evaluate the contribution of the integral called quasi-moduli integral(QMI). It is shown that this can be evaluated explicitly by the calculation using exact WKB.
We also show that the partition function itself is invariant under the Borel sum.

\subsubsection{Comparison to quasi-moduli integral}
%
Using the decomposition of resolvent given in \eqref{resolvent-1}, we can 
write  the partition function  as
\begin{align}
Z             & =Z_{\rm p}(\beta)+Z_{\rm np}(\beta)                                                                                                        
\end{align}
where 
\begin{align}
	Z_{\rm p}(\beta)    & =\frac{1}{2\pi i}\int_{\epsilon-i\infty}^{\epsilon+i\infty}\qty[-\pdv{E}\log (1+A)]e^{-\beta E} dE+\frac{1}{2\pi i}\int_{\epsilon-i\infty}^{\epsilon+i\infty}\qty[-\pdv{E}\log (1+A^{-1})]e^{-\beta E} dE,                                  
\end{align}
and
\begin{align}
	Z_{\rm np}(\beta) & =\frac{1}{2\pi i}\int_{\epsilon-i\infty}^{\epsilon+i\infty}\qty[-\pdv{E}\log\qty(1+\frac{B}{D_A^2})]e^{-\beta E} dE                 \nonumber\\
	              & =-\beta\frac{1}{2\pi i}\int_{\epsilon-i\infty}^{\epsilon+i\infty}\log\qty(1+\frac{B}{D_A^2})e^{-\beta E} dE                         \nonumber\\
                & =\beta\frac{1}{2\pi i}\int_{\epsilon-i\infty}^{\epsilon+i\infty}\sum_{n=1}^\infty\frac{1}{n}\qty(-\frac{B}{D_A^2})^ne^{-\beta E} dE \,,
\end{align}
where we have used integration by parts moving to the second line. 
We now clarify the relation between the above quasi-moduli integral and our result on the non-perturbative contribution
\begin{align}
	Z_{\rm np}(\beta)=\beta\frac{1}{2\pi i}\int_{\epsilon-i\infty}^{\epsilon+i\infty}\sum_{n=1}^\infty\frac{1}{n}\qty(-\frac{B}{D_A^2})^ne^{-\beta E} dE \,,
\end{align}
Using this asymptotic expansion in Eq.~(\ref{Asymptotic_omega}) $A\sim e^{-2\pi i \frac{E}{\hbar \omega_A(E,\hbar)}}$ again. Then $D_{A}$ is given by
\begin{align}
	D_{A}=\begin{cases}
    e^{\pi i\frac{E}{\hbar\omega_A}}2\sin\qty(\pi\qty(\frac{E}{\hbar\omega_A}+\frac{1}{2}))=e^{\pi i\frac{E}{\hbar\omega_A}}\frac{2\pi}{\Gamma(\frac{1}{2}+\frac{E}{\hbar\omega_A})\Gamma(\frac{1}{2}-\frac{E}{\hbar\omega_A})} & (\Im \hbar > 0)\\
    e^{-\pi i\frac{E}{\hbar\omega_A}}2\sin\qty(\pi\qty(\frac{E}{\hbar\omega_A}+\frac{1}{2}))=e^{-\pi i\frac{E}{\hbar\omega_A}}\frac{2\pi}{\Gamma(\frac{1}{2}+\frac{E}{\hbar\omega_A})\Gamma(\frac{1}{2}-\frac{E}{\hbar\omega_A})} & (\Im \hbar < 0)\,.
  \end{cases}
\end{align}
For our purpose, we drop the irrelevant Gamma function factor $\frac{\sqrt{2\pi}}{\Gamma(\frac{1}{2}+\frac{E}{\hbar\omega_A})}$, which corresponds to the negative eigenvalue and does not contributes to the integral in the case of harmonic oscillator.  

Then, we rewrite $Z_{\rm np}$ as
\begin{align}
	Z_{\rm np}(\beta)=\beta\frac{1}{2\pi i}\int_{\epsilon-i\infty}^{\epsilon+i\infty}\sum_{n=1}^\infty\frac{1}{n}\qty(-1)^n\qty(B\frac{e^{\mp 2\pi i \frac{E}{\hbar\omega_A}}}{4\sin^2\qty(\pi\qty(\frac{E}{\hbar\omega_A}+\frac{1}{2}))})^n e^{-\beta E} dE\,.
\end{align}

By defining $s\equiv E/(\hbar \omega_{A})-1/2$,
it is expressed as
\begin{align}
    Z_{\rm np}(\beta)&=\beta\frac{1}{2\pi i}\int_{\epsilon-i\infty}^{\epsilon+i\infty}\sum_{n=1}^\infty\frac{1}{n}(-1)^n\qty(B\frac{e^{\mp 2\pi i (\frac{1}{2}+s)}}{4\sin^2(\pi s)})^n e^{-\beta (\hbar\omega_A(\frac{1}{2}+s))} \hbar\omega_A ds 
\end{align}
Essentially $\frac{\sqrt{2\pi}}{\Gamma(1+s)}$ in $2\sin(\pi s)=\frac{2\pi}{\Gamma(-s)\Gamma(1+s)}$ corresponds to the negative eigenvalues, so if we define the integral path to take only positive eigenvalues, this integral can be approximated\footnote{The residues of $\frac{1}{\sin(\pi s)}$ and $\Gamma(-s)$ are different, so just removing $\Gamma(1+s)$, even though it essentially corresponds to the negative eigenvalues, would change the result. However if we only consider the residue around $s=0$, which corresponds to the ground state energy, the factor $\Gamma(1+s)$ can be ignored.} as 
\begin{align}
  Z_{\rm np}(\beta)&\simeq\beta\frac{1}{2\pi i}\int_{\epsilon-i\infty}^{\epsilon+i\infty}\sum_{n=1}^\infty\frac{1}{n}(-1)^n\qty(\Gamma\qty(-s)^2\frac{B}{2\pi}e^{\mp 2\pi i (1/2+s)})^n e^{-\beta (\hbar\omega_A(1/2+s))} \hbar\omega_A ds    \label{QMIform_Maslov} \\
  &=\beta\frac{1}{2\pi i}\int_{\epsilon-i\infty}^{\epsilon+i\infty}\sum_{n=1}^\infty\frac{1}{n}\qty(B\Gamma\qty(-s)^2\frac{1}{2\pi}e^{\mp 2\pi i s})^n e^{-\beta\frac{\hbar\omega_A}{2}}e^{-s\beta} \hbar\omega_A ds\,.
  \label{QMIform_final}
\end{align}
Here, the partition function obtained by calculating the path integral is as follows\cite{Sueishi}\cite{ZinnJustin:2004ib}, (Appendix \ref{Sueishi_doublewell}):
\begin{align}
  \frac{Z_{\rm np}}{Z_0}=\beta\frac{1}{2\pi i}\int_{\epsilon-i\infty}^{\epsilon+i\infty}\sum_{n=1}^\infty\frac{1}{n}\qty(e^{-S_{\rm bion}}\qty(\frac{\det M_{I}}{\det M_0})^{-1}\frac{S_{\rm inst}}{2\pi}\Gamma\qty(-s)^2\qty(\frac{\hbar}{2})^{-s}e^{\mp 2\pi i s})^n e^{-s\beta} ds\,.
\end{align}
Comparing this with our results, we can see that we obtain the path integral representation and indeed each sector in Eq.~(\ref{QMIform_final}) has physical meaning as follows: $\beta$ corresponds to the zero-mode integral (translation symmetry of time-dependent solution), $\frac{1}{n}$ is cyclic permutation of multi-bions, $\Gamma(-s)^2e^{\mp 2\pi i s}$ are quasi-moduli integrals(QMI) (See Appendix \ref{sec:QMI_explanation}) with Stokes phenomena\footnote{There are two QMI for one bion and it gives two Gamma functions.}, $B=e^{-\frac{1}{\hbar}\oint_B p}+O(\hbar)$ is the bion contribution with quantum correction, the integral from $-i\infty$ to $i\infty$ corresponds to the delta function constraint in the quasi-moduli integral, and $e^{-\frac{1}{2}\beta\hbar\omega_A}$ is regarded as the partition function of perturbative part in the large $\beta$ limit,
$Z_0$. The missing part in Eq.~(\ref{QMIform_final}) is $S_{\rm inst}\qty(\frac{\det M_{I}}{\det M_0})^{-1}$ and $\qty(\frac{\hbar}{2})^{-s}$. However, both are coming from the quantum fluctuation. But since the quantum fluctuation is included in $B$ and $\omega_A$, it is considered to be required by doing a higher-order expansion of them.
$(-1)^n$ in Eq.~(\ref{QMIform_Maslov}) is regarded as the \textit{Maslov index}. The origin of this index is easily understood by using Gutzwiller's quantization as shown in Sec.~\ref{sec:Gutzwiller_quantization}. This factor is cancelled by $e^{\mp \pi i n}$ in Eq.~(\ref{QMIform_Maslov}), which looks like the hidden topological angle (HTA) \cite{Behtash:2015loa}) of $n$-bion configuration though, the index has very important role: this quantity can be regarded as the intersection number of Lefschetz thimble\footnote{The relation between the Maslov index and Lefschetz thimbles is also discussed in \cite{Behtash:2017rqj}}.

Furthermore, comparing the QMI calculation and Gutzwiller's perspective, we can see the new physical meaning of QMI. The QMI calculation is based on the approximation that the cycle is sufficiently large, but from Gutzwiller's point of view, the $B$ cycle is so short that it requires the $A$ cycle to rotate infinite times in order to earn the sufficiently long cycle, and therefore it is considered to be represented in the form of $D_A^{-2}B$. This perspective explains the riddle in the calculation of \cite{Sueishi}\cite{ZinnJustin:1981dx}: The form of the $\Gamma$ function derived from the vacuum and the form of the $\Gamma$ function derived from QMI matched despite both were calculated entirely separately. 
\begin{align}
  D(E)=\frac{1}{\Gamma\qty(\frac{1}{2}-E)\Gamma\qty(\frac{1}{2}-E)}\qty(1-Be^{\pm i\pi\qty(1-2E)}\qty(\frac{\hbar}{2})^{\qty(1-2E)}\Gamma\qty(\frac{1}{2}-E)\Gamma\qty(\frac{1}{2}-E))=0
  \label{conspire}
\end{align}
The first $\frac{1}{\Gamma\qty(\frac{1}{2}-E)\Gamma\qty(\frac{1}{2}-E)}$ are from two vacua and the latter ones are from QMI. This miracle is easily explained by this Gutzwiller's representation. Both have essentially the same origin, the infinite number of $A$ cycles, $D_A^{-1}=\frac{1}{1+A}
=\sum_{n}^\infty(-1)^nA^n$. 

\subsubsection{The intersection number of Lefschetz thimble}

It is notable that the of quantization condition in Eqs.~(\ref{eq:D_hbp}) determines the ``relevant saddles" in the path integral and the intersection number of Lefschetz thimble ($n_\sigma$). Firstly, as we mentioned in Sec.~\ref{sec:PI}, the Fredholm determinant can be expressed as
\begin{align}
	D(E)=\prod_\sigma D_\sigma^{n_\sigma}(E)\,.
\end{align}
Now, the quantization condition given by Eq.~(\ref{eq:D_hbp}) can be rewritten as
\begin{align}
  D&=(1+A)(1+A^{-1})\qty(1+\frac{B}{D_A^2})\nl
  &=(1+A)(1+A^{-1})\qty[ e^{-\frac{B}{D_A^2}}]^{-1}\qty[e^{-\frac{1}{2}\qty(\frac{B}{D_A^2})^2}]\qty[e^{-\frac{1}{3}\qty(\frac{B}{D_A^2})^3}]^{-1}\qty[e^{-\frac{1}{4}\qty(\frac{B}{D_A^2})^4}]...\nl
  &=(1+A)(1+A^{-1})\prod_{n=1}^\infty D_n^{(-1)^n}
  \label{Maslov_thimble_doublewell}
\end{align}
The first $(1+A)$ and $1+A^{-1}$ are regarded as the Fredholm determinant coming from the vacuum saddle points and the latter ones are ones from $n$-bion saddle points. 
\begin{align}
    D_n&=e^{-\frac{1}{n}\qty(\frac{B}{D_A^2})^n} \label{eq:Dn_Lef} \\
    Z_n&=\frac{1}{2\pi i}\int_{-i\infty}^{i\infty}\qty[\pdv{E}\frac{1}{n}\qty(\frac{B}{D_A^2})^n]e^{-\beta E}dE \nonumber\\
    &=\frac{\beta}{2\pi i}\int_{\epsilon-i\infty}^{\epsilon+i\infty}\frac{1}{n}\qty(\Gamma\qty(-s)^2\frac{B}{2\pi}e^{\mp 2\pi i(1/2+s)})^n e^{-\beta (\hbar\omega_A(1/2+s))} \hbar\omega_A ds\,. \label{eq:Zn_Lef}
\end{align}
This representation is a factorized form for each bion, and as explained in the previous section, the $Z_n$ obtained from this $D_n$ is indeed the partition function of n-bions. The power of each bracket $[...]$ is nothing other than the Maslov index. Therefore, we can see the intersection number of Lefschetz thimble of non-perturbative contributions is exactly corresponding to the Maslov index. 

Strictly speaking, the Maslov index is attached to both $A$ and $B$ cycles. However, the former is not regarded as the intersection number of Lefschetz thimble but the only latter's is. This difference is related to the following situation:
In the case of harmonic oscillator, we often calculate the partition function around the vacuum, and indeed it gives the correct answer. However there are other classical solutions in this system s.t. time-dependent solutions like oscillating around the vacuum. If we choose such the solution as the saddle point, we still get the same partition function. The reason is such time-dependent solutions are included in quantum fluctuations around the vacuum. On the other hand, the non-perturbative saddle, bions should be summed up to obtain the correct partition function. It means $n$-bion (and the vacuum)is topologically separated in the functional space, or one  can say $2$-bion cannot be expressed as $1$-bion with quantum fluctuation.

We would make some comments on the Maslov index and intersection number.
We may also write the quantization conditions for the symmetric double well as follows:
\begin{align}
    D&=(1+A)(1+A^{-1})\qty(1+\frac{B}{D_A^2})\nl
    &=(1+A)(1+A^{-1})\sqrt{1+\frac{B}{D_A^2}}\sqrt{1+\frac{B}{D_A^2}}\nl
    &=(1+A)(1+A^{-1})\qty(\prod_{n=1}^\infty D_n^{\frac{1}{2}(-1)^n})\qty(\prod_{n=1}^\infty D_n^{\frac{1}{2}(-1)^n}), \label{eq:D_split}
\end{align}
and this form can be considered as giving a fractional intersection number, $\mp \frac{1}{2}$. 
In the similar way to the procedure for obtaining the partition function given by Eq.(\ref{eq:Zn_Lef}), the quantization condition (\ref{eq:D_split}) gives
\begin{align}
    Z&\simeq Z_{1,{\rm pert}}+Z_{2,{pert}}+\frac{1}{2}\sum_{n=1}^\infty e^{-nS_{\rm bion}}+\frac{1}{2}\sum_{n=1}^\infty e^{-nS_{\rm bion}}.
\end{align}
This form corresponds to the representation of the vacuum with a different starting point as a separate term. However, while physically it is reasonable to write the contributions in this way, from the point of view of transseries, these term should be combined.

Also the following form is considerable.
\begin{align}
    D&=(1+A)(1+A^{-1})\qty(1+\frac{B}{D_A^2})\nl
    &=(1+A)(1+A^{-1})\qty(1+i\frac{\sqrt{B}}{D_A})\qty(1-i\frac{\sqrt{B}}{D_A})
\end{align}
The latter parts corresponds to the instanton contributions. Because of $G_{inst.}(E)=-\pdv{E}\log(1+i\frac{\sqrt{B}}{D_A})\propto\sum_{n}(-i)^n \qty(\frac{\sqrt{B}}{D_A})^n$, the Maslov index of instanton ($\sqrt{B}$) is $-i$.  
Using parity operator, $\hat{P}\ket{x}=\ket{-x}$, we can consider the projected partition function $Z_{\pm}=\tr\qty(\frac{1\pm \hat{P}}{2}e^{-\beta \hat{H}})$. From \cite{ZinnJustin:1981dx}, it corresponds to
\begin{align}
    Z_+&\rightarrow(1+A)\qty(1+i\frac{\sqrt{B}}{D_A})\nl
    Z_-&\rightarrow(1+A^{-1})\qty(1-i\frac{\sqrt{B}}{D_A})
\end{align}
Therefore, if we impose the non-periodic boundary condition on the path integral, such the noninteger intersection number can appear. However, if we consider only the periodic trajectory, the Maslov index is always integer, which means the intersection number is also integer.


\subsection{Delabaere-Dillinger-Pham (DDP) formula}
\label{sec:DDP}
In this subsection, we would like to briefly review the Delabaere-Dillinger-Pham (DDP) formula\cite{DDP1}.
For the double-well potential, as we have seen the previous sections, the $A,C$-cycles are defined as  asymptotic series. These series are  non-Borel nonsummable  if ${\rm Im} \, \hbar = 0$. 
When the an asymptotic expansion is  Borel nonsummable for  ${\rm Im} \, \hbar = 0$,  the Borel transformed cycles have a singular point on the positive real axis of the Borel plane, in other words an imaginary ambiguity happens according to the choice of the sign of ${\rm Arg} (\hbar)$ for the Borel resummation.

By employing this imaginary ambiguity the information of $A,C$-cycles can be carried into the $B$-cycle via the Stokes automorphism.
This relationship is so called \textit{the resurgence relation}. This type of resurgence relation connects high orders of the asymptotic expansion of the  $A,C$-cycles to low orders perturbative expansion of $B$-cycle. 
In the physical sense, the $A,C$-cycles and $B$-cycle are now interpreted as a perturbative expansion(fluctuation) in terms of $\hbar$ around (locally) bounded potential and the nonperturbative bion background, respectively, so that their asymptotic expansions can be related to each others by the resurgence relation.
Instead of directly looking at the Borel plane, there exits a way to find the same relation from the Stokes graph of the exact WKB analysis, which is so called \textit{the Delabaere-Dillinger-Pham (DDP) formula}.
The DDP formula can be directly applied to any functions of the cycles, and it would also have the important role to see the cancellation of imaginary ambiguities for the partition function, discussed in the later section.
From here, we would like to  demonstrate some physical applications to (DDP) formula  \cite{DDP1,Iwaki1} to potential-well problems. 

\begin{figure}[t]
	\centering 
	\includegraphics[width=8cm]{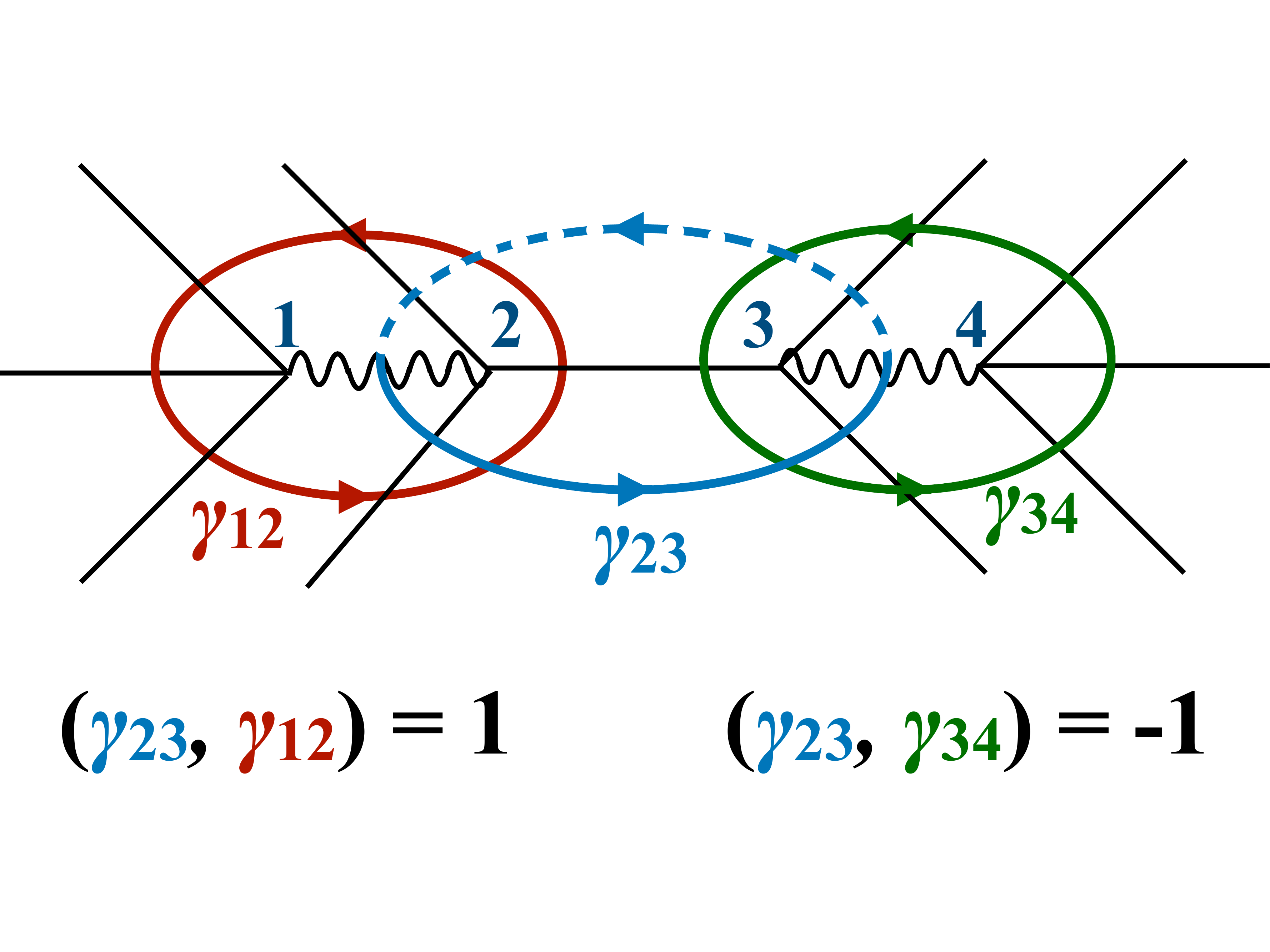}
	\caption{
          Intersection number among cycles in the DDP formula.
          In this case, the intersection numbers of cycles are given by $(\gamma_{23},\gamma_{12})=+1$ and $(\gamma_{23},\gamma_{34})=-1$.
          (See fig.\ref{fig:DDP1} for the definition.)
        }
        \label{fig:DDP2}
\end{figure}

\begin{figure}[t]
	\centering 
	\includegraphics[width=8cm]{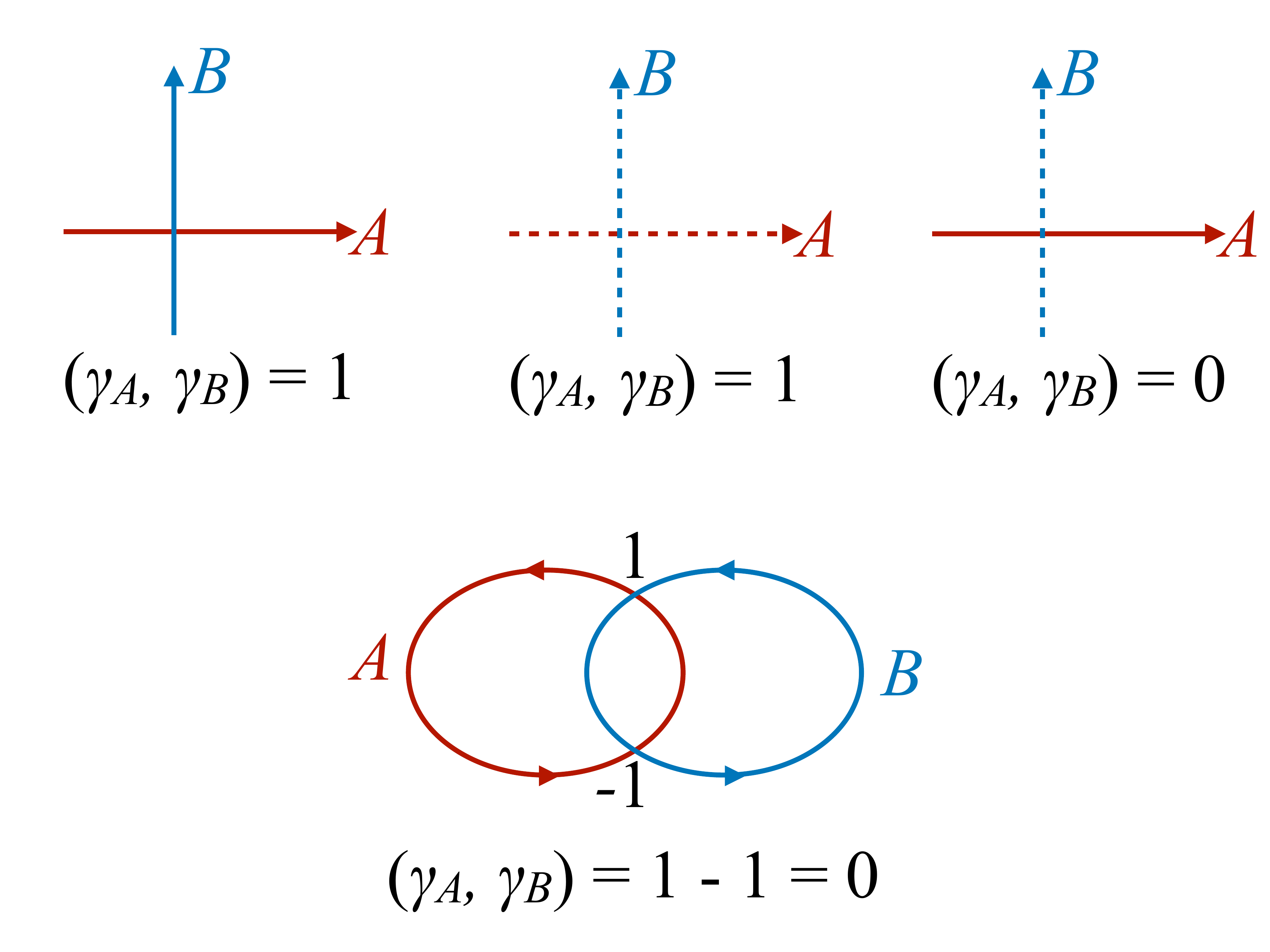}
	\caption{
          The definition of intersection number.
          The solid and dashed lines denote a part of cycle on the first and second Riemann sheet, respectively.
          If two solid(dashed) lines given by $\gamma_A$ and $\gamma_B$ intersect with each other in respect to a right-handed coordinate, we say $(\gamma_A,\gamma_B)=+1$.
          If the intersection is given by lines on different sheets from each other, we say $(\gamma_A,\gamma_B)=0$.
          For example, two cycles not crossing branch-cuts always give $(\gamma_A,\gamma_B)=0$.
        }
        \label{fig:DDP1}
\end{figure}

For simplicity, let us start with the simple setup which is the double-well potential with low energy shown in Fig.~\ref{fig:DDP2}.
The Stokes graph has four turning points on the real axis labelled by $a_1< \cdots < a_4$, and we consider three anti-clockwise cycles enclosing a pair of two turning points. These are $(a_1,a_2)$, $(a_2,a_3)$, $(a_3,a_4)$.
We named these oriented loops as $\gamma_{12},\gamma_{23},\gamma_{34}$ in Fig.\ref{fig:DDP2}, and the quantities  $A,B,C$-cycles are defined along the each loops.
Since the branch-cut lay on the complex $x$-plane with the endpoints at turning points, the $A$ and $C$-cycles are now defined on the first Riemann have nontrivial value.
In contrast, the $B$-cycle twice crosses the independently defined branch-cut, and the lower(upper) half contour lies on the first(second) Riemann sheet.
This means that the $B$-cycle has the intersection with each $A$- and $C$-cycles once on the first sheet but does not on the second sheet\footnote{If the $A (C)$-cycle is defined on the second sheet, the relation with the cycle on the first sheet can be found as 
\begin{align}
A_{\rm 1st. sheet} = 1/A_{\rm 2nd. sheet}.
\end{align}
}.
Under this setup, the DDP formula is obtained as follows. Here all symbols such as $A$, $B$ used so far means Borel summed ones. In order to make this point clear, we will describe its asymptotic form as $\tilde{A}$, $\tilde{B}$ and Borel summed ones as $\mathcal{S}_\pm[\tilde{A}]$ and $\mathcal{S}_\pm[\tilde{B}]$ in this sectiopn. The DDP formula is given as
\begin{align}
	\mathcal{S}_+[\tilde{A}] & =\mathcal{S}_-[\tilde{A}] (1+\mathcal{S}[\tilde{B}])^{-1}, \label{eq:DDP_A}\\ 
	\mathcal{S}_+[\tilde{B}] & =\mathcal{S}_-[\tilde{B}] =: \mathcal{S}[\tilde{B}],         \\
	\mathcal{S}_+[\tilde{C}] & =\mathcal{S}_-[\tilde{C}] (1+\mathcal{S}[\tilde{B}])^{+1}, \label{eq:DDP_C}
\end{align}
where ${\cal S}_{\pm}$ is the Borel resummation for ${\rm sign} ({\rm Im} \, \hbar) = \pm 1$.
The exponent of $(1+\mathcal{S}[\tilde{B}])$ is by intersection number $(\gamma_A,\gamma_B) = \pm 1$ which is defined as follows:
If the intersection between perturbative and non-perturbative cycles, $A$ and $B$, is right(left)-handed, we say that $(\gamma_A,\gamma_B)=+1(-1)$.
If a perturbative cycle does not have intersection with non-perturbative cycles, then it gives $(\gamma_A,\gamma_B)=0$.
Fig.~\ref{fig:DDP1} shows how to determine the intersection number.

In the previous section we separately obtained the quantization condition $D$ for ${\rm Im}\,\hbar >0$ and ${\rm Im}\,\hbar <0$, but one can see that these can be related to each other via the DDP formula:
\begin{align}
  {\cal S}_{+}[\tilde{D}^+] =  {\cal S}_{-}[\tilde{D}^{-}] , \label{eq:DDP_SD}
\end{align}
where $D^{+}$ and $D^{-}$ are given by eq.(\ref{eq:D_hbp}) for the positive and negative ${\rm Im} \hbar$, respectively, but replaced $A,B,C$ with $A^\pm,B,C^\pm$.
It is important to mention that from eq.(\ref{eq:DDP_SD}) one can indeed show the imaginary ambiguity cancellation for \textit{all order of bion sectors.}
We will summarize the proof in Appendix \ref{sec:Im_cancel}.

\subsubsection{Unambiguity  of the partition function under the Borel resummation}
\label{sec:path-integ}

In this section, we will show that the partition function itself does not have Borel ambiguity, that is, the partition function itself does not change due to the sign of the imaginary part of the $\hbar$. To show this statement, we now rewrite the quantization condition in Eq.~(\ref{eq:D_hbp}) in a way that clearly shows the imaginary amgibuity part as 
 \be
 \mathcal{S}_\pm[A]=\hat{A}\pm\delta \hat{A}.
 \ee
  Then we can show the quantization condition can be expressed as the form without the imaginary ambiguity term (See Appendix \ref{sec:Im_cancel}.)
We now have
\be
\lim_{{\rm Arg}(\hbar) \rightarrow 0_{\pm}} {\cal S}_{\pm}[\tilde{D}^{\pm}/\Omega(\tilde{L})]
&=&  (1 + \hat{A}) (1 + \hat{C}) +  \frac{\hat{B}}{2+\hat{B}} \left( \hat{A} + \hat{C} \right) - \frac {\hat{B}^2}{(\hat{B}+2)^2} \hat{A}\hat{C},
\label{D_invariant}
\ee
where  the symbol $\hat{\frak C}$ is called medianization and is defined as: 
\begin{align}
\hat{\frak C}:=( \lim_{\hbar \rightarrow 0_+}{\cal S}_{+}[\tilde{\frak C}] + \lim_{\hbar \rightarrow 0_-}{\cal S}_{-} [\tilde{\frak C}])/2, \qquad  {\frak C} \in \{A,B,C\}
\end{align}
The $A$- and $C$-cycles are related to each other, $\hat{A}=1/\hat{C}=\hat{C}^*$. ${\cal S}_{\pm}(\Omega(L))=-({A}{B}{C} )^{-1}\ne 0$ is an invariant quantity under the lateral Borel summation $\mathcal{S}_\pm$, which does not contribute to the partition function itself.
\be 
&&  G(E)  =-\pdv{E}\log {\cal S}_{+}[\tilde{D}^{+}/\Omega(\tilde{L})]  =G_{\rm p}(E)+G_{\rm np}(E),  \\
&&  G_{\rm p}(E)  = -\pdv{E}\left[ \log  ( 1+\hat{A} ) + \log ( 1+\hat{C} ) \right], \\
&&  G_{\rm np}(E)  = -\pdv{E}\log \left[1+ \frac{ \hat{A}  + \hat{C} }{( 1+\hat{A} ) ( 1+\hat{C} )} 
\cdot \frac{\hat{B}}{2+\hat{B}} - \frac{ \hat{A} \hat{C} }{( 1+\hat{A} ) ( 1+\hat{C})} \cdot \frac{\hat{B}^2}{(2+\hat{B})^2}\right], \label{eq:Gnp}
\ee
Therefore, the partition function can be expressed by
\be
&& Z(\beta)  = Z_{\rm p}(\beta)+Z_{\rm np}(\beta), \\
&& Z_{\rm p}(\beta)  = -\frac{1}{2\pi i}\int_{\epsilon-i\infty}^{\epsilon+i\infty} G_{\rm p}(E) e^{-\beta E}  dE, \\
&& Z_{\rm np}(\beta) = -\frac{1}{2\pi i}\int_{\epsilon-i\infty}^{\epsilon+i\infty} G_{\rm np}(E) e^{-\beta E}  dE.
\ee
It is notable that the $n$-th bion sector can be identified from Eq.~(\ref{eq:Gnp}) by taking a form that 
\be
G_{\rm np}(E) = -\frac{\pd}{\pd E} \sum_{n=1}^{\infty} \hat{B}^{n} \Xi^{(n)}(\hat{A}),
\ee
where
\be
\Xi^{(n)}(\hat{A}) 
&=& \frac{(-1)^{n+1}}{2^n} \sum_{k=1}^{\infty} \sum_{m=k}^{2k} \frac{1}{k}
\begin{pmatrix}
 k \\
 m-k
\end{pmatrix}
 \frac{\left( \hat{A}+\hat{A}^{-1}  \right)^{2k-m}}{(\hat{D}_{{A}}^{+} \hat{D}_{{A}}^{-})^k}
 \sum_{{\bf s} \in {\mathbb N}_0^{\otimes m-1}}^{|{\bf s}|=n-m} \frac{(n-m)!}{s_1 ! \cdots s_{m-1} ! \left( n- m - |{\bf s}| \right)!},
\ee
where $|\bullet|$ denotes the 1-norm and we assumed that $ \sum_{{\bf s} \in {\mathbb N}_0^{\otimes m-1}}^{|{\bf s}|=n-m} \cdots = 1$ when $m=1$.
Here, we used $\hat{D}_{{A}}^{\pm} =1+\hat{A}^{\pm 1}$ and  $\hat{A}=\hat{C}^{-1}$.

By construction, Eq.~(\ref{D_invariant}) is invariant under lateral Borel summation so this form of the partition function is also invariant\footnote{Of course this invariance can be checked with DDP formula.} under $\mathcal{S}_{\pm}$. 
\section{Generic potentials ($N$ple-well potential)} 
\label{sec:Generic}

The procedure to derive the quantization condition from the exact-WKB analysis and the construction of the path-integral explained in Sec.~\ref{sec:DW} can be extended to the cases with more generic forms of potentials.
Since the extension is straightforward, we would show only the quantization condition and make some comments.

\begin{figure}
	\centering 
	\includegraphics[width=15cm]{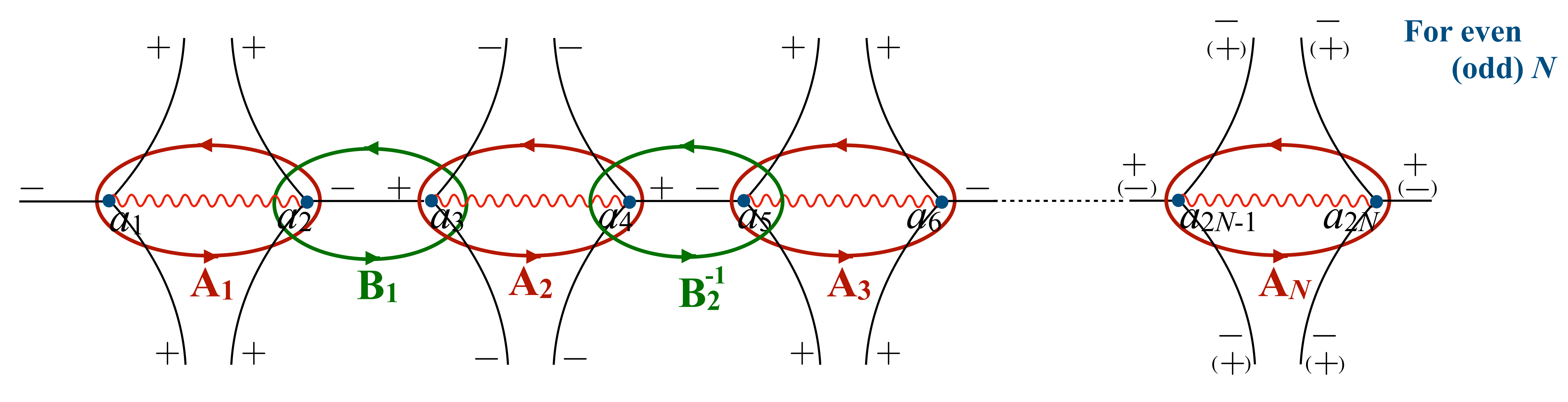}
	\caption{
	The Stokes graph for the $N$ple-well potential for ${\rm Arg}(\hbar)=0$.
	Perturbative cycles $A_n$ and nonperturbative cycles $B_n$ are defined as cycles enclosing a branch-cut and a stokes line having end points at turning points, respectively.
	}
        \label{fig:DDP-N}
\end{figure}

We now focus on a parity-symmetric real-bounded potential,  $V(x) \in {\mathbb R}[[x^2]]$ with $\lim_{|x|\rightarrow \infty}V(x) = + \infty$. There are $N$ perturbative and $N-1$  non-perturbative cycles as shown in Fig.~\ref{fig:DDP-N}.
By repeating the same argument as that in Sec.~\ref{sec:DDP}, the DDP formula can be easily extended to these cases.
When the $N$ non-perturbative cycles ($\tilde{B}_i$) lays on the complex $x$-plane, the DDP formula for a  perturbative cycle ($\tilde{A}_{j}$) can be expressed as \cite{DDP1}
\begin{align}
  \mathcal{S}_+[\tilde{A}_j] & =   \mathcal{S}_-[\tilde{A}_j] \prod_{i=1}^{N -1} (1+\mathcal{S}[\tilde{B}_i])^{(-1)^{i} \cdot (\gamma_{{B_i}},\gamma_{A_j})}\,, \\
   \mathcal{S}_+[\tilde{B}_i] & = \mathcal{S}_-[\tilde{B}_i] =: \mathcal{S}[\tilde{B}_i] \quad \mbox{for all} \quad  1 \le i \le N\,.
\end{align}
In this notation, we chose the orientation of $\tilde{B}_i$-cycles  such that $\tilde{B}_{i}$ is exponentially small.
Then, by considering the path-orbit of the wave function from the left to the right and using the connection formula, the quantization condition for the positive/negative ${\rm Arg}(\hbar)$ can be obtained as
\be
&& \tilde{D}^{+}_{N} = \Omega(\tilde{L}) \sum_{{\bf n} = {\bf 0}}^{|{\bf n}|=N} \;  \prod_{k = 1}^{\lfloor (N+1)/2 \rfloor} \left[ \left( 1 + \tilde{B}_{{2k-2}} \right)^{1-n_{2k-2}} \left( 1+ \tilde{B}_{{2k-1}} \right)^{1-n_{2k}} \right]^{n_{2k-1}} \cdot \Phi^{({\bf n})}(\tilde{\bf A}), \label{Nple_Dp}\\
&& \tilde{D}^{-}_{N} = \Omega(\tilde{L})  \sum_{{\bf n} = {\bf 0}}^{|{\bf n}|=N} \;  \prod_{k = 1}^{\lfloor N/2 \rfloor}
\left[ \left( 1 + \tilde{B}_{{2k-1}} \right)^{1-n_{2k-1}} \left( 1+ \tilde{B}_{{2k}} \right)^{1-n_{2k+1}} \right]^{n_{2k}} \cdot \Phi^{({\bf n})}(\tilde{\bf A}), \label{Nple_Dm}\\
&& \Omega(\tilde{L}) = \frac{i^{N}}{\tilde{L}}, \qquad \tilde{L}:= \left( \prod_{k=1}^{N} \tilde{A}_{k} \right) \left( \prod_{k=1}^{N-1} \tilde{B}_{k} \right), 
\ee
where ${\bf n} =(n_0,n_1,\cdots,n_N,n_{N+1}) \in ({\mathbb Z}/2 {\mathbb Z})^{\otimes N+2}$ with $n_0 = n_{N+1}=0$, $|{\bf n}|$ denotes the 1-norm of ${\bf n}$, $\tilde{B}_{0}=\tilde{B}_{N}=0$, and $\Phi^{({\bf n})}(\tilde{\bf A}):= \prod_{k=1}^{N} \tilde{A}^{n_k}_k$.\footnote{For $\tilde{D}^-_{N=1}$, we assume that $\prod_{k=1}^{0} \left[ \cdots \right]=1$ .}
In the similar way to the double-well potential summarized in Appendix \ref{sec:Im_cancel}, one finds ${\cal S}_{+}[\tilde{D}^+]={\cal S}_{-}[\tilde{D}^-]$ and shows that {\it the imaginary ambiguity is cancelled for all order of bion sectors.} 
After the cancellation,  the quantization condition can be expressed in terms of unambiguous parts as: 
\be 
&& \lim_{{\rm Arg}(\hbar) \rightarrow 0_{\pm}} {\cal  S}_{\pm}[\tilde{D}^{\pm}_{N}/\Omega(\tilde{L})] \nl
&=& \sum_{{\bf n} = {\bf 0}}^{|{\bf n}|=N} \frac{2 \bar{\frak S}_0^{({\bf n})}(\hat{\bf B})}{\bar{\frak S}_0^{({\bf n})}(\hat{\bf B}) + 1} \cdot \prod_{k=1}^{\lfloor(N+1)/2\rfloor} \left[ \left( 1 + \hat{B}_{{2k-2}} \right)^{1-n_{2k-2}} \left( 1+ \hat{B}_{{2k-1}} \right)^{1-n_{2k}} \right]^{n_{2k-1}} \cdot \Phi^{({\bf n})}(\hat{\bf A}), \nl
&=& \sum_{{\bf n} = {\bf 0}}^{|{\bf n}|=N} \frac{2}{\bar{\frak S}_0^{({\bf n})}(\hat{\bf B}) + 1} \cdot \prod_{k=1}^{\lfloor N/2 \rfloor} \left[ \left( 1 + \hat{B}_{{2k-1}} \right)^{1-n_{2k-1}} \left( 1+ \hat{B}_{{2k}} \right)^{1-n_{2k+1}} \right]^{n_{2k}}  \cdot \Phi^{({\bf n})}(\hat{\bf A}), \nl
&=& \sum_{{\bf n} = {\bf 0}}^{|{\bf n}|=N}   \frac{2 \prod_{k=1}^{N}[(1+\hat{B}_{k-1})(1+\hat{B}_{k})]^{n_{k}} \cdot \Phi^{({\bf n})}(\hat{\bf A})}{\prod_{k=1}^{\lfloor(N+1)/2\rfloor}[(1+\hat{B}_{2k-2})^{1+n_{2k-2}}(1+\hat{B}_{2k-1})^{1+n_{2k}}]^{n_{2k-1}} + \prod_{k=1}^{\lfloor N/2 \rfloor}[(1+\hat{B}_{2k-1})^{1+n_{2k-1}}(1+\hat{B}_{2k})^{1+n_{2k+1}}]^{n_{2k}}}, \nl
\label{eq:tilD_N}
\ee
where
\be
&& \Phi^{({\bf n})}(\hat{\bf A}) = \frac{\lim_{{\rm Arg}(\hbar)  \rightarrow 0_+}{\cal S}_+[\Phi(\tilde{\bf A})]+ \lim_{{\rm Arg}(\hbar)  \rightarrow 0_-} {\cal S}_-[\Phi(\tilde{\bf A})]}{2}, \\
&& \hat{B}_{n}:=
\begin{cases}
  \lim_{{\rm Arg}(\hbar) \rightarrow 0_{+}}{\cal S}_+ [\tilde{B}_{n}] = \lim_{{\rm Arg}(\hbar) \rightarrow 0_{-}}{\cal S}_- [\tilde{B}_{n}] & \quad \mbox{for} \quad  1 \le n < N \\
 0 & \quad \mbox{for} \quad  n = 0, N 
\end{cases},
\ee
  $\bar{\frak S}^{({\bf n})}_0(\hat{\bf B})$ is given by ${\cal S}_{+} [\Phi^{({\bf n})}(\hat{\bf A})] = {\cal S}_{-} \circ {\frak S}_{0}[\Phi^{({\bf n})}(\hat{\bf A})]=: \bar{\frak S}^{({\bf n})}_0(\hat{\bf B}) \Phi^{({\bf n})}(\hat{\bf A})$ with the Stokes automorphism being expressed as ${\frak S}_{0}$.
Notice that $\Omega(L)$ does not contribute to the path-integral.
Because of parity symmetry and the reality condition for the potential, the different cycles have nontrivial relation

\be
\hat{A}_{n} = 
\begin{cases}
\hat{A}^{-1}_{N-n+1} = [\hat{A}_{N-n+1}]^* & \mbox{for even } N \\
\hat{A}_{N-n+1} = [\hat{A}_{N-n+1}^{-1}]^* & \mbox{for odd } N
\end{cases}, \qquad
\hat{B}_{n} = \hat{B}_{N-n} = [\hat{B}_{N-n}]^*.
\ee

Thus, eq.(\ref{eq:tilD_N}) is real for an even $N$.
For an odd $N$, the real $D^{\pm}_{N}$ can be obtained by dividing by $\hat{A}_{(N+1)/2}^{1/2} \cdot \prod_{n=1}^{(N-1)/2} \hat{A}_{n}$, which does not contribute to calculation of residue for reproducing the partition function due to the Cauchy's argument principle.

One can find the origin of the intersection number (index) of the thimble decomposition, i.e. the Maslov index, in terms of the quantization condition.
For example, the quantization condition (\ref{Nple_Dp}), (\ref{Nple_Dm}) for $N=3$ can be expressed as

\begin{align}
D^+_{N=3}/\Omega(L) 
& = (1 + A_1)(1 + A_2)(1 + A_1) \left[ 1 +  \frac{B}{ D_{A_1}^{-2} D_{A_2}^{+}} \left(  2 D_{A_1}^{-}  + B \right) \right] \,
,   \\
D^-_{N=3}/\Omega(L) 
& =  (1 + A_1) (1 + A_2)(1 + A_1) \left[ 1 + \frac{B}{D_{A_1}^{+2} D_{A_2}^{-} } \left( 2 D_{A_1}^{+}  + B   \right) \right]\,.
\end{align}

where ${\frak C}:=\lim_{{\rm Arg}(\hbar) \rightarrow 0_{+(-)}} {\cal S}_{+(-)}[\tilde{\frak C}]$ for cycles ${\frak C} \in \{A_1,A_2,A_3,B_1,B_2 \}$ in $D^{+(-)}_{N=3}$, and 
\begin{align}
D_{A_1}^{\pm} := 1 + A_{1}^{\pm 1}, \qquad D_{A_2}^{\pm} := 1 + A_{2}^{\pm 1}\,.
\end{align}
We also use the fact that $A_3 = A_1$ and $B_1 = B_2 =: B$.
Therefore, we obtain

\begin{align} 
 D^\pm_{N=3}/\Omega(L) 
 =&  (1+A_1) ( 1+A_2 )(1+A_1)  \nl
 & \cdot \prod_{n=1}^{\infty} \exp \left[  -\frac{1}{n}  \left\{  \frac{2 B} {D_{A_1}^{\mp} D_{A_2}^{\pm}}  + \frac{B^2}{D_{A_1}^{\mp 2} D_{A_2}^{\pm}} \right\}^n \right]^{(-1)^{n}}\,.
\end{align}

As shown in Eq.~(\ref{Maslov_thimble_doublewell}), the power $(-1)^n$ is the Maslov index of each nonperturbative cycle and is regarded as the intersection number of Lefschetz thimble. In the similar way to Eqs.~(\ref{eq:Dn_Lef})-(\ref{eq:Zn_Lef}), the partition function in the asymptotic limit can be expressed as
\begin{align}
    Z=Z_{\rm p}+Z_{\rm np}\,,
\end{align}
with
\begin{align}
    Z_{\rm p}&=\frac{1}{2\pi i}\int_{\epsilon-i\infty}^{\epsilon+i\infty} \qty[-\pdv{E}\log (1+A_1)]e^{-\beta E} dE \nl
    & \ +\frac{1}{2\pi i}\int_{\epsilon-i\infty}^{\epsilon+i\infty}\qty[-\pdv{E}\log (1+A_2)]e^{-\beta E} dE+\frac{1}{2\pi i}\int_{\epsilon-i\infty}^{\epsilon+i\infty}\qty[-\pdv{E}\log (1+A_1)]e^{-\beta E} dE\,,\nl 
\end{align}
    \begin{align}
    Z_{\rm np}&=\frac{\beta}{2\pi i}
    \int_{\epsilon-i\infty}^{\epsilon+i\infty}\sum_{n=1}^\infty\frac{1}{n}(-1)^n\qty[  \frac{2 B} {D_{A_1}^{\mp} D_{A_2}^{\pm}}  + \frac{B^2}{D_{A_1}^{\mp 2} D_{A_2}^{\pm}} ]^n \nl
    &\simeq\frac{\beta}{2\pi i}
    \int_{\epsilon-i\infty}^{\epsilon+i\infty}\sum_{n=1}^\infty\frac{1}{n}(-1)^n  
    \left[
    \left. e^{\mp \pi i\qty(\frac{E}{\hbar\omega_{A_1}}+\frac{E}{\hbar\omega_{A_2}})}
    \frac{B}{2\pi}
    \Gamma \left( \frac{1}{2}-\frac{E}{\hbar\omega_{A_1}} \right)
    \Gamma \left( \frac{1}{2}-\frac{E}{\hbar\omega_{A_2}} \right) 
    \right. \right.  \nl
& \ \left.    
    +e^{\mp \pi i\qty(2 \frac{E}{\hbar\omega_{A_1}}+\frac{E}{\hbar\omega_{A_2}})}
    \frac{B^2}{(2\pi)^{3/2}}\Gamma \left(\frac{1}{2}-\frac{E}{\hbar\omega_{A_1}} \right)^2
    \Gamma \left(\frac{1}{2}-\frac{E}{\hbar\omega_{A_2}} \right)
    \right]^n e^{-\beta E}dE\,,    \label{eq:triple_Z}
\end{align}

\begin{figure}[t]
	\centering 
	\includegraphics[width=12cm]{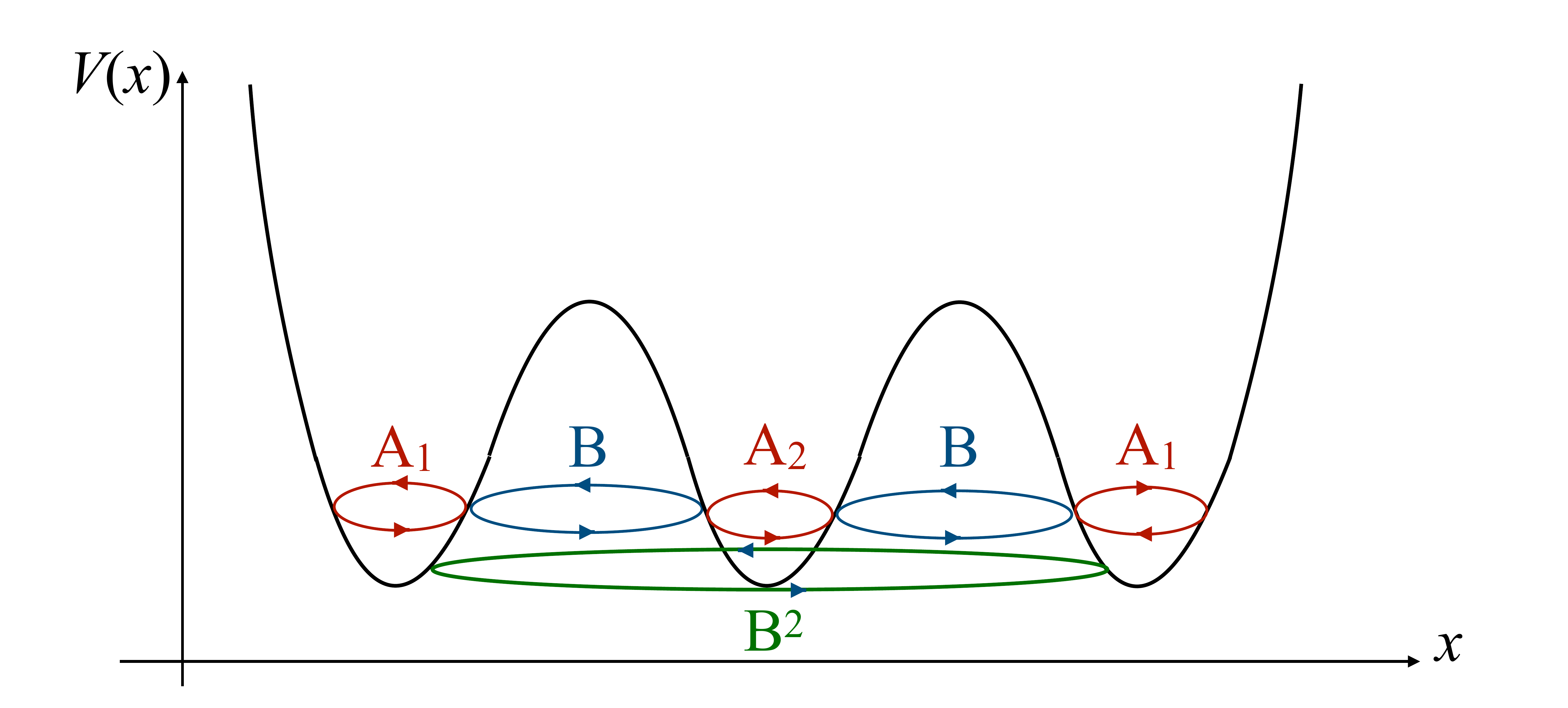}
	\caption{   
	Relationship between periodic orbits and the Maslov index for the symmetric triple-well potential.
	The index $(-1)^n$ in the $n$-the sector in Eq.(\ref{eq:triple_Z}) is determined by counting a cycle-unit which can be decomposed into the following parts; two $B/(D_{A_1} D_{A_2})$ and one $B^2/(D_{A_1}^2 D_{A_2})$.
	The former part including one $B$-cycle corresponds to an orbit running around two (locally) double-well potential,  [$A_1$-$B$-$A_2$] and [$A_2$-$B$-$A_1^{-1}$], whereas the latter part including two $B$-cycle is related to an orbit which is globally running from the left to the right [$A_1$-$B$-$A_2$-$B$-$A_1^{-1}$].
	}
        \label{fig:Triple_cycle}
\end{figure}
where $\simeq$ means dropping the factor $\frac{\sqrt{2\pi}}{\Gamma(\frac{1}{2}+\frac{E}{\hbar\omega_A})}$, essentially the negative eigenvalue part. Now, the non-perturbative contribution of the partition function for the triple-well potential can be interpreted as follows. There are two types of bions as shown in Fig.~\ref{fig:Triple_cycle}: the one that reverses immediately at the turning point (blue) and the other passing through the barrier (green). The former gives $B\frac{1}{D_{A_1}D_{A_2}}\sim e^{-S_{bion}}\Gamma(\frac{1}{2}-\frac{E}{\hbar\omega_{A_1}})\Gamma(\frac{1}{2}-\frac{E}{\hbar\omega_{A_2}})$ and the latter gives $B^2\frac{1}{D_{A_1}D_{A_2}D_{A_1}}\sim e^{-2S_{bion}}\Gamma(\frac{1}{2}-\frac{E}{\hbar\omega_{A_1}})^2\Gamma(\frac{1}{2}-\frac{E}{\hbar\omega_{A_2}})$.
We can identify how to sum up the non-perturbative contributions and this result shows the structure of partition function as a trans-series\footnote{The two blue bion and one green bion produce the same amplitude $B^2\sim e^{-2S_{bion}}$ but these should be distinguished. This is quite natural from the path integral view because these are different classical solutions.}. The phase ambiguity is interpreted as the reversal of the direction of the cycle $A_2$. We also note the Maslov index is not completely cancelled like the case of double-well potential because $e^{\mp \pi i\qty(2 \frac{E}{\hbar\omega_{A_1}}+\frac{E}{\hbar\omega_{A_2}})}$ produces $(\mp i)^n$, not $(-1)^n$. 

The results in this section clearly indicate that the findings in the paper are generic and applicable at least to
the symmetric multi-well potential quantum mechanics.
The extension to the other systems is also possible, but it is  left for future work.



\section{Discussion and Summary}
\label{sec:summary}
In this work, we explored the connections between exact WKB method, saddle point analysis of Euclidean path integration, and the Gutzwiller trace formula. Our main findings can be listed as follows: 

\begin{enumerate}

\item Both exact WKB and saddle point method applied to path integrals take  place in complex domain. In exact WKB, position $x$ is complexified, and  in path integrals, the space of paths is  complexified.   Stokes phenomena permeate through   both constructions. 
We showed  that  the Stokes phenomena in exact-WKB  expressed through exact quantization condition   maps to the Stokes phenomena in the saddle-point method in path integration.  

  \item  Traditionally, the instanton methods are commonly used to deduce ground state (or few  excited state) properties by considering  $\beta \rightarrow \infty$ limit.   If we wish to derive the instanton contribution to arbitrary energy level, we need to study 
  the partition function at finite $\beta$. At this end, 
  we found a new  physical interpretation  of the quasi-moduli integral (QMI) in the semiclassical analysis of path integral, which relates it to 
  the fluctuation determinant around  the harmonic minima.    For the perturbative saddle, we are considering fluctuation operator  around $x_{\rm cr} (\tau)=0$, which ultimately produce  harmonic state sums
  $\sum_{n=0}^{\infty} e^{-\beta n (\omega+ \frac{1}{2})}$ with the perturbative corrections.  The QMI at finite-$\beta$ for the bion configuration also produce the same factor, see \ref{conspire}.  Despite apparent difference,  this is physically reasonable, because the bion configurations is also almost        $x_{\rm cr} (\tau)=0$ everywhere, except for the instanton and anti-instanton cores,  and this  is   natural from the  perspective of Gutzwiller's quantization condition.  
  This conspiracy has to take place in order the Euclidean path integral to reproduce    correctly the level number dependence 
  over the Hilbert space, and we believe  this is important for future work.

   \item
The Maslov index that  appears in the non-perturbative contribution  is identified with the intersection number of 
  corresponding Lefschetz thimble.

  \item 
We showed the equivalence of Bohr-Sommerfeld, Gutzwiller, path-integral quantization conditions via the exact-WKB analysis. 
  Exact-WKB  and Stokes graphs are  extremely powerful tools, and we believe that there is much to be learned by using them. 
  
  \item 
The resurgent structures of partition functions naturally continues to hold  for generic symmetric multi-well potential, and ambiguity cancellations 
holds to all orders in semi-classical expansion. We showed this by using exact-WKB.
	      
\end{enumerate}


Our results not only uncover the unknown facts on quantization conditions and resurgent structures in quantum mechanics but also exhibit that the exact-WKB method and the Stokes curves can be powerful tools to study physical problems.
Below, we describe a couple of examples to which we hope to  apply our methods

\begin{itemize}

\item In semi-classics approach to Euclidean path integral formulation,  the action must be complexified at the beginning  to  determine the set of saddles that can possibly contribute. 
 But  it is not always easy to know which complex saddles contribute and which do not.  
Exact WKB method, via the use of basic Stokes graph, immediately answers this part of question.  Furthermore, it can also be used to determine the Stokes multipliers of the corresponding saddles. There is  clearly much to be learned in semi-classical  approach to path integral by using the knowledge of  exact (complex) WKB.

	\item 
	A streamlined construction of the semi-classical expansion of Euclidean partition function  at finite $\beta$,  which is capable of addressing all the states in the Hilbert space,  addressing both below  and above the barrier  in the spectrum, and
	not just restricted to few   lowest lying states as usually done in literature. 
	
	  \item    Analysis of graded partition functions in quantum mechanical systems  coupled to Grassmann valued fields (e.g. supersymmetric or quasi-exactly solvable systems) or equivalently, Wess-Zumino terms, i.e, path integral of a particle $x(\tau)$ with intrinsic spin. We hope to discuss both path integral as well as   the 	exact-WKB construction for these systems. 
    	      	      
	\item 
	      Quantum mechanics of particles on $S^1$ in the presence of a potentials  $-\cos(N x), x \sim x+ 2\pi $ and topological theta angle.  Study of such systems using exact WKB methods would be interesting. 
	    
    	      	      
	\item
	Detailed investigation of  classical and quantum tilted multi-well systems by using exact-WKB. 
    	      	      
    	      	      
	\item An  outstanding question is whether the constructive  (low-order/low order) resurgence between P/NP sectors e.g. \cite{Dunne:2013ada, Alvarez3},  (e.g. connecting pertubation theory around perturbative vacuum to perturbation theory around instanton)	 
	can be generalized to systems with higher genus with the help of exact WKB formula.

	\item For the application to QFT, we can apply our method to simplified QFT with approximations or reductions giving 1D QM systems, such as $S^1$ compactification from 2D QFTs and integrable systems. (Schwinger model in two dimensions, ${\mathbb C}P^{N}$-model, Toda lattice, etc.)
	Furthermore, since the Gutzwiller trace formula itself can be applied to QFTs under semiclassical approximation \cite{DHN}, we can consider the hidden relation between the Maslov index and the intersection number of Lefschetz thimbles in the similar fashion.
\end{itemize}

\begin{acknowledgements}
    S. \ K. and N. \ S. especially thank A. Behtash for grateful discussions of connection between the exact WKB analysis and Lefschetz thimble formulation.
    N. \ S. and T.\ M. appreciate Kohei Iwaki for giving useful information and detailed explanations about exact WKB analysis.  
	T.\ M. is supported by the Japan Society for the Promotion of Science (JSPS) Grant-in-Aid for Scientific Research (KAKENHI) Grant Numbers 18H01217 and 19K03817.
	The original questions related to the present work were posed in ``RIMS-iTHEMS International Workshop on Resurgence Theory" 
    at RIKEN, Kobe in 2017. The authors are grateful to the organizers and participants of the workshop.
	M.~U. acknowledges support from U.S. Department of Energy, Office of Science, Office of Nuclear Physics under Award Number DE-FG02-03ER41260.
\end{acknowledgements}


\appendix

\section{Quasi-moduli integral(QMI)}
\label{sec:QMI_explanation}

We now give a brief review on the quasi-moduli integral in quantum mechanics.
One of the methodologies for studying nonperturbative aspects of quantum theory in the Euclidean setup is the dilute instanton gas approximation (DIGA), where we ignore the interaction between the instantons.
However, the recent study on quantum mechanics and low-dimensional quantum field theories uncovers that DIGA is not sufficient in terms of the nontrivial relation, called resurgent relation, between large-order growth of perturbative series and nonperturbative contribution from instanton--anti-instanton pair (bion) \cite{Dunne:2012ae,Dunne:2012zk,Misumi:2014jua,Misumi:2014bsa,Misumi:2015dua,Fujimori:2016ljw,Fujimori:2017oab,Fujimori:2017osz,Fujimori:2018kqp}. Here, both of the perturbative Borel resummation and the nonperturbative bion contribution are accompanied with imaginary ambiguities, which are cancelled out. 
This bion amplitude is calculated via quasi-moduli integral (QMI) of the interaction potential between instanton and anti-instanton. Quasi-moduli parameters are not genuine moduli but come to be moduli in the well-separated limit of the distance of instanton and anti-instanton \cite{Behtash:2015zha,Fujimori:2016ljw,Fujimori:2017oab,Fujimori:2017osz,Fujimori:2018kqp}.

One of the properties of the quasi-moduli integral is that the results includes gamma functions.
Below, we show this fact briefly.
For instance, the effective potential for a bion in the double-well quantum mechanics is given by
\begin{align}
V(\tau) \,=\, 
{1\over{3\hbar}}-{2\over{\hbar}}\,e^{-\tau}\, + \epsilon \tau,
\end{align}
where $\tau$ stands for the distance between instanton and anti-instanton, which is nothing but a quasi moduli in this case. 
The last term $\epsilon \tau$ corresponds to the deformation term originating in quantum-mechanical fermionic degrees of freedom. We are interested only in the bosonic quantum mechanics, thus we take a $\epsilon\to 0$ limit in the end of calculation.
It is shown in Ref.~\cite{Behtash:2015zha} that this integral is performed by complexifying $\tau$ and applying the Lefschetz thimble decomposition of the integration contour, which corresponds to the thimble decomposition of the complexified path integral associated with complex saddle points.
We instead calculate it in a distinct but equivalent manner.
In this method, since the integral is not convergent in a small $\hbar$ region, we have to take a prescription called Bogomol'nyi--Zinn-Justin prescription~\cite{ZinnJustin:1981dx}, where we first regard $-\hbar$ as positive-valued and take analytic continuation as $-\hbar = e^{\pm i \pi} \hbar$ in the end. 
The bion contribution to the partition function is calculated via the quasi-moduli integral as
\begin{align}
QMI = {\pi \hbar \over{\beta e^{-2S_{I}}}}{Z_{I\bar{I}}\over{Z_{0}}} &\,=\,e^{2S_{I}} \int_{-\infty}^{\infty} d\tau e^{-V(\tau)}
\nonumber\\
&\,=\,  \int_{-\infty}^{\infty} d\tau \exp \left[-{2\over{-\hbar}} e^{-\tau} - \epsilon \tau
\right]
\nonumber\\
&\,=\, \int_{0}^{\infty} {ds\over{s}} e^{-s} \left( {-\hbar\over{2}} s \right)^{\epsilon}\quad\quad\quad\quad\quad \left(s\equiv {2\over{-\hbar}} e^{-\tau}\right)
\nonumber\\
&\,=\, \Gamma(\epsilon) \left( {-\hbar\over{2}} \right)^{\epsilon}.
\end{align}
where $S_{I} = 1/(3\hbar)$ is the instanton action and $\beta$ is an imaginary time period.
One finds a gamma function emerges in this result.
By taking analytic continuation as $-\hbar=\hbar e^{\pm i \pi}$, where the sign corresnponds to the direction of its analytic continuation, and also it is regarded as the Stokes phenomena of this integral. The symmetric double well corresponds to taking $\epsilon\to 0$ limit but its gives $1/\epsilon$ pole, which is known as $1/\epsilon$ problem.

In Ref.~\cite{Sueishi}, the above calculation of the quasi-moduli integral for double-well quantum mechanics is extended to the $n$-bion contribution, which is obtained as
\begin{align}
  QMI^n &= \frac{1}{n}\prod_{i=1}^{2n}\qty(\int_0^\infty d\tau_i e^{-\mathcal{V}_i(\tau_i)})\delta\qty(\sum_{k=1}^{2n} \tau_k-\beta) \nl
  &=\frac{1}{n}\prod_{i=1}^{2n}\qty(\int_0^\infty d\tau_i e^{-\mathcal{V}_i(\tau_i)})\frac{1}{2\pi}\int_{-\infty}^\infty dl e^{il\sum_{k=1}^{2n} (\tau_k-\beta)} \nl
  &=\frac{1}{n}\frac{1}{2\pi i}\int_{-i\infty}^{i\infty}ds e^{-s\beta}\qty(\int_0^\infty d\tau e^{(s-\epsilon)\tau+\frac{2}{\hbar}e^{-\tau}})^{2n}\,.
\end{align}
We use the following relation,
\begin{align}
  \int_0^\infty d\tau e^{(s-\epsilon)\tau+\frac{2}{\hbar}e^{-\tau}}=e^{\pm i\pi (\epsilon-s)}\qty(\frac{\hbar}{2})^{\epsilon-s}\Gamma(\epsilon-s)\,,
\end{align}
Here the $\pm$ is corresponded to the sign of $\Im(\hbar)$, which comes from Stokes phenomenon of this quasi-moduli integral. the symmetric double well 
then, we obtain the final expression as
\begin{align}
  QMI^n(\epsilon\rightarrow0)=\frac{1}{2\pi i n}\int_{-i\infty}^{i\infty}dse^{-s\beta}\qty(e^{\pm i\pi (-s)}\qty(\frac{\hbar}{2})^{-s}\Gamma(-s))^{2n}
  \label{eq:QMI}
\end{align}
The partition function of symmetric double well is this form:
\begin{align}
  \frac{Z}{Z_0}&=2\qty(1 + \sum_{n=1}^\infty \beta B^n QMI^{n}(0))
  \label{Sueishi_doublewell}
\end{align}
 The two $Z_0$ is from the two vacua\footnote{If we do the calculation with $\epsilon$ finite, we need to take into account the differences of the energy of each vacuum and QMI.See~\cite{Sueishi}} and $B$ is a bion contribution (with its fluctuation). The rest of summation is from multi-bion and the linear $\beta$ is from the translation symmetry of (imaginary) time dependent solutions.

If we set $\mathcal{V}_i(\tau_i)=0$, which means the dilute gas approximation, the QMI becomes
\begin{align}
  \qty(\prod_{i=1}^{2n}\int_0^\infty d\tau_i) \delta\qty(\sum_{k=1}^{2n} \tau_k-\beta)=\frac{1}{(2n-1)!}\beta^{2n-1}
\end{align}
Then the partition function is
\begin{align}
  \frac{Z}{Z_0}=2\sum_{n=0}^{\infty}\frac{B^n\beta^{2n}}{(2n)!}=2\cosh(\sqrt{B}\beta)
  \label{DIGbion}
\end{align}
This is usual dilute instanton gas approximation of symmetric double well potential.

\section{Imaginary ambiguity cancellation for the double well potential} 
\label{sec:Im_cancel}

In this appendix, we would show that for the symmetric double-well potential
  \be 
  \lim_{{\rm Arg}(\hbar) \rightarrow 0_{+}} {\rm Im}\, {\cal S}_{+}[\tilde{D}^+/\Omega(\tilde{L})] = \lim_{{\rm Arg}(\hbar) \rightarrow 0_{-}} {\rm Im}\, {\cal S}_{-}[\tilde{D}^-/\Omega(\tilde{L})],    \label{eq:ImSD0}
  \ee
  where ${\cal S}_{\pm}[\tilde{D}^{\pm}]$ is given by eq.(\ref{eq:DDP_SD}).
  
  In order to avoid confusion, we would like to employ the following notation for any cycles ${\frak C}(\hbar) \in \{ A(\hbar),B(\hbar),C(\hbar) \}$:
  \be
    \tilde{\frak C}(\hbar) &:& \mbox{Asymptotic expansion or transseries, e.g. $\sum_{n} a_{n}\hbar^n$ or $e^{-c/\hbar}\sum_{n} b_{n}\hbar^n$}, \nl
  {{\frak C}}^{\pm}(\hbar) &:& \mbox{Borel resummation of $\tilde{\frak C}(\hbar)$}, \nn
  \ee
  where we take ${\rm Arg}(\hbar) = 0$ and define ${\frak C}^{\pm}(\hbar)$ as 
  \be
    {\frak C}^{\pm}(\hbar) := 
    \lim_{{\rm Arg}(\hbar) \rightarrow 0_{\pm}} {\cal S}_{\pm} [\tilde{\frak C}] (\hbar). 
  \ee
  Since the $A$- and $C$-cycles are Borel nonsummable when ${\rm Arg}(\hbar) = 0$, one finds that 
  \be
  {A}^{+}(\hbar) \ne {A}^{-}(\hbar), \qquad {C}^{+}(\hbar) \ne {C}^{-}(\hbar), \qquad {B}^{+}(\hbar) = {B}^{-}(\hbar) =: {B}(\hbar), 
  \ee
  because of the Borel singularity, i.e., the imaginary ambiguity.
  If we define 
  \be 
  \hat{\frak C}(\hbar):= \frac{{\frak C}^{+}(\hbar) + {\frak C}^{-}(\hbar)}{2},\qquad
  \delta \hat{\frak C}(\hbar):= \frac{{\frak C}^{+}(\hbar) - {\frak C}^{-}(\hbar)}{2},
  \ee
  then $\hat{\frak C}(\hbar) \sim \tilde{\frak C} (\hbar)$ for any cycles and
  the ambiguity $\delta \hat{\frak C}(\hbar)$ can be evaluated from the DDP formula, (\ref{eq:DDP_A})-(\ref{eq:DDP_C}).
  Since the DDP formula gives
  \be
  && 
  \left\{
  \begin{alignedat}{4}
    {A}^{+}(\hbar)  &=& {A}^{-}(\hbar)  \left( 1 + {B}^{-}(\hbar) \right)^{-1} \\
    {C}^{+}(\hbar)  &=& {C}^{-}(\hbar)  \left( 1 + {B}^{-}(\hbar) \right)^{+1} 
  \end{alignedat} 
  \right. \nl \nl
  &\Rightarrow \qquad&
  \left\{
  \begin{alignedat}{4}
    \hat{A}(\hbar) + \delta \hat{A}(\hbar)  &=&  \left( \hat{A}(\hbar) - \delta \hat{A}(\hbar) \right) \left( 1 + \hat{B}(\hbar) \right)^{-1} \\
    \hat{C}(\hbar) + \delta \hat{C}(\hbar)  &=&  \left( \hat{C}(\hbar) - \delta \hat{C}(\hbar) \right) \left( 1 + \hat{B}(\hbar) \right)^{+1} 
  \end{alignedat}
  \right.,
  \ee
  one can find that
  \be
  \delta \hat{A}(\hbar) = - \frac{\hat{B}(\hbar)}{2 + \hat{B}(\hbar)} \cdot \hat{A}(\hbar), \qquad \delta \hat{C}(\hbar) = + \frac{\hat{B}(\hbar)}{2 + \hat{B}(\hbar)} \cdot \hat{C}(\hbar).
  \ee
  For a monomial basis in terms of cycles, $\Phi(\tilde{A},\tilde{B},\tilde{C})$, its imaginary ambiguity can be computed as
  \be
  {\Phi}(\hat{A},\hat{B},\hat{C}) &:=& \frac{1}{2}  
  \left( \lim_{\theta \rightarrow 0_+}{\cal S}_+ + \lim_{\theta \rightarrow 0_-} {\cal S}_- \right) [\Phi(\tilde{A},\tilde{B},\tilde{C})]
  \\
  \delta {\Phi}(\hat{A},\hat{B},\hat{C}) &:=& \frac{1}{2} \left( \lim_{\theta \rightarrow 0_+} {\cal S}_+ -  \lim_{\theta \rightarrow 0_-} {\cal S}_- \right) [\Phi(\tilde{A},\tilde{B},\tilde{C})]
  = \frac{\bar{\frak G}_{0}({B})-1}{\bar{\frak G}_{0}({B})+1}\Phi(\hat{A},\hat{B},\hat{C}),
  \ee
  where $\bar{\frak G}_{0}(\bar{B})$ is a function of $\tilde{B}$ satisfying
  \be
     {\cal S}_{+}[\Phi(\tilde{A},\tilde{B},\tilde{C})] = {\cal S}_{-} \circ {\frak G}_{0} [\Phi(\tilde{A},\tilde{B},\tilde{C})] = \bar{\frak G}_{0}(\hat{B}) {\cal S}_{-}[\Phi(\tilde{A},\tilde{B},\tilde{C})],
  \ee
  with the Stokes automorphism ${\frak G}_{0}$.
  One can check that
  \be
  {\Phi}(\hat{A},\hat{B},\hat{C}) + \delta {\Phi}(\hat{A},\hat{B},\hat{C}) &=&  \frac{2 \bar{\frak G}_{0}(\hat{B})}{\bar{\frak G}_{0}(\hat{B})+1}\Phi(\hat{A},\hat{B},\hat{C}) \nl
  &=&  \bar{\frak G}_{0}(\hat{B}) \left[{\Phi}(\hat{A},\hat{B},\hat{C}) - \delta {\Phi}(\hat{A},\hat{B},\hat{C}) \right].
  \ee
  For example, if one takes $\Phi(\tilde{A},\tilde{B},\tilde{C}) = \tilde{L} := \tilde{A}\tilde{B}\tilde{C}$, the ambiguity is zero because ${\frak G}_{0}[\tilde{L}]=\tilde{L}$ ($\bar{\frak G}_{0}(\hat{B})=1$).

  We would rewrite the quantization condition in terms of ${A}(\hbar)$ and ${C}(\hbar)$.
  From the explicit calculation, one can make sure that the DDP formula for the quantization condition (\ref{eq:DDP_SD}) is indeed satisfied and it can be written down as
  \be
  \lim_{{\rm Arg}(\hbar) \rightarrow 0_{+}} \, {\cal S}_{+}[\tilde{D}^+/\Omega(\tilde{L})] 
  &=& (1 + \hat{A}(\hbar)) (1 + \hat{C}(\hbar)) +  \frac{\hat{B}(\hbar)}{2+\hat{B}(\hbar)} \left( \hat{A}(\hbar) + \hat{C}(\hbar) \right) \nl 
  &=& \lim_{{\rm Arg}(\hbar) \rightarrow 0_{-}} \, {\cal S}_{-}[\tilde{D}^-/\Omega(\tilde{L})]. \label{eq:qcond_explicit}
  \ee
  
  Then, we would try to make the relationship between $A(\hbar)$ and $C(\hbar)$.
  From the definition,
  \be
  \hat{A}(\hbar) = \exp \left( \oint_{\gamma_{12}} dx \, S_{\rm odd}(x,\hbar)\right),\qquad \hat{C}(\hbar) = \exp \left( \oint_{\gamma_{34}} dx \, S_{\rm odd}(x,\hbar)\right).
  \ee
  Notice that $a_{1}=-a_{4}$ and $a_{2}=-a_{3}$.
  Hence, the contour integration can be written as
  \be
  \oint_{\gamma_{12}} dx \, S_{\rm odd}(x,\hbar) &=& \int_{a_1-\delta_{1}-\delta_2}^{a_2 +\delta_1- \delta_2} dx \, S_{\rm odd}(x,\hbar) + \int_{a_2 +\delta_1- \delta_2}^{a_2+\delta_{1}+\delta_2} dx \, S_{\rm odd}(x,\hbar) \nl
  && + \int_{a_2+\delta_{1}+\delta_2}^{a_1 -\delta_1+ \delta_2} dx \, S_{\rm odd}(x,\hbar) + \int_{a_1 -\delta_1 + \delta_2}^{a_1-\delta_{1}-\delta_2} dx \, S_{\rm odd}(x,\hbar) \nl
  &=& -\int_{a_4+\delta_{1}+\delta_2}^{a_3 -\delta_1 + \delta_2} dx \, S_{\rm odd}(-x,\hbar) - \int_{a_3 -\delta_1+ \delta_2}^{a_3-\delta_{1}-\delta_2} dx \, S_{\rm odd}(-x,\hbar) \nl
  && - \int_{a_3-\delta_{1}-\delta_2}^{a_4 +\delta_1- \delta_2} dx \, S_{\rm odd}(-x,\hbar) - \int_{a_4 +\delta_1 - \delta_2}^{a_4+\delta_{1}+\delta_2} dx \, S_{\rm odd}(-x,\hbar) \nl
  &=& - \oint_{\gamma_{34}} dx \, S_{\rm odd}(-x,\hbar), \label{eq:gm12_gm34}
  \ee
  where $0 < \delta _{1,2} \ll 1 \in {\mathbb R}_{+}$.
  Since $S_{\rm odd}(x,\hbar)=S_{\rm odd}(-x,\hbar)$ for the symmetric double-well potential, one finds that $\hat{A}(\hbar) = 1/\hat{C}(\hbar)$.\footnote{
  This condition depends on the choice of branchcuts and the orientation of cycles.
  In our case we assume that the cuts are taken as defined in Fig.\ref{fig:DDP-N}.
  For a generic $N$-well potential based on Fig.\ref{fig:DDP-N}, the parity symmetry for $S_{\rm odd}(x)$ acts as $S_{\rm odd}(x,\hbar)=(-1)^{N+1} S_{\rm odd}(-x,\hbar)$, and thus the reality condition is also modified as $\hat{A}_{n} = (-1)^N [\hat{A}_{N-n+1}]^*$ with $1 \le n \le N$.}
  In addition, the complex conjugation for $A$-cycle gives $[\hat{A}(\hbar)]^*=1/\hat{A}(\hbar)$, which implies that $[\hat{A}(\hbar)]^*= \hat{C}(\hbar)$.
  Therefore, eq.(\ref{eq:qcond_explicit}) is real, which gives the statement (\ref{eq:ImSD0}).
  
  Notice that this discussion is applicable to any bounded potentials preserving parity symmetry and reality condition.

\end{document}